\documentclass[journal]{IEEEtran}
\usepackage{xcolor}
\usepackage{steinmetz}

\RequirePackage{xcolor} 

\RequirePackage[cmex10,intlimits]{amsmath}
\RequirePackage{amsfonts}
\RequirePackage{amssymb}

\providecommand{\J}{\ensuremath{\mathrm{j}}}    
\providecommand{\RE}{\ensuremath{\mathrm{Re}}}  
\providecommand{\IM}{\ensuremath{\mathrm{Im}}}  

\providecommand{\Quot}[1]{``{#1}"}              
\providecommand{\D}{\,\mathrm{d}}               
\providecommand{\V}[1]{\boldsymbol{#1}}         
\providecommand{\M}[1]{\mathbf{#1}}             
\providecommand{\T}[1]{\mathrm{#1}}             
\providecommand{\UV}[1]{\V{\hat{#1}}}           
\providecommand{\OP}[1]{{\mathcal{#1}}}         

\providecommand{\herm}{\mathrm{H}}

\providecommand{\ZVAC}{\ensuremath{Z_0}}           

\providecommand{\basisFcn}{\V{\psi}}

\providecommand{\Ivec}{\ensuremath{\M{I}}}
\providecommand{\Vvec}{\ensuremath{\M{V}}}
\providecommand{\Fvec}{\ensuremath{\M{K}}}

\providecommand{\Iopt}{\Ivec_\T{ub}}
\providecommand{\Gopt}{G_\T{ub}}
\providecommand{\Goptsr}{G_\T{ub}^\T{sr}}

\providecommand{\Lmax}{L_\T{max}}
\providecommand{\IVEC}{[\Ivec_n]}

\providecommand{\Rmat}{\ensuremath{\M{R}}}
\providecommand{\Xmat}{\ensuremath{\M{X}}}
\providecommand{\Zmat}{\ensuremath{\M{Z}}}

\providecommand{\Lmat}{\M{L}}

\providecommand{\Prad}{P_\T{rad}}

\providecommand{\Plost}{P_\T{lost}}
\providecommand{\Ohm}{\Omega}

\newcommand{\ie}{\textit{i}.\textit{e}.{}} 
\newcommand{\eg}{\textit{e}.\textit{g}.{}}
\newcommand{\cf}{\textit{cf}.{}}

\usepackage[]{hyperref}
\hypersetup{
    colorlinks,
    linkcolor={red!50!black}, 
    citecolor={blue!50!black},
    urlcolor={blue!20!black}
}

\usepackage{graphicx}
\graphicspath{{figures/}}

\begin{document}
\title{The Upper Bound on Antenna Gain and Its Feasibility as a Sum of Characteristic Gains}
\author{Miloslav~Capek, \IEEEmembership{Senior Member, IEEE} and Lukas~Jelinek
\thanks{Manuscript received \today; revised \today. This work was supported by the Czech Science Foundation under project~\mbox{No.~21-19025M} and by the Ministry of Education,
Youth and Sports under Project LTAIN19047.}
\thanks{M. Capek and L. Jelinek are with the Czech Technical University in Prague, Prague, Czech Republic (e-mails: \{miloslav.capek; lukas.jelinek\}@fel.cvut.cz).}
\thanks{Color versions of one or more of the figures in this paper are
available online at http://ieeexplore.ieee.org.}
}

\maketitle

\begin{abstract}
The upper bound on antenna gain is expressed as a sum of lossy characteristic modes, specifically, as a sum of characteristic far fields squared. The procedure combines the favorable properties of Harrington's classical approach to maximum directivity and current-density-based approaches. The upper bound is valid for any antenna or array designed in a given design region for which optimal performance is determined. The decomposition into modes makes it possible to study the degrees of freedom of an obstacle, classify its radiation into normal or super-directive currents, and determine their compatibility with a given excitation. The bound considers an arbitrary shape of the design region and specific material distribution. The cost in Q-factor and radiation efficiency is studied. The extra constraint of a self-resonance current is imposed for an electrically small antenna. The examples verify the developed theory, demonstrate the procedure's utility, and provide helpful insight to antenna designers. The feasibility of the optimal gain is studied in detail on an example of end-fire arrays using the aforementioned decomposition of optimal current density into lossy characteristic modes.
\end{abstract}


\section{Introduction}
\label{Sec:Intro}

\IEEEPARstart{A}{ntennas} and antenna arrays with high-gain performance are required in many applications~\cite{Ziolkowski_MixturesOfMultipoles}. Their design is intricate, especially in the case of an electrically small design region~\cite{Fujimoto_Morishita_ModernSmallAntennas, VolakisChenFujimoto_SmallAntennas} or high-gain end-fire arrays~\cite{Yaghjian_etal_RS2008, Jaafar2018, Karasawa2022, Debard2023}. Knowing the theoretical limits at the beginning of the design procedure is, therefore, advantageous since it allows us to judge the performance of realized designs~\cite{Tang_etal_CompactLowProfileLinCircPolFiltennas, Chen_etal_WidebandHighDensityCircularlyPolarizedArray, Gaponenko2021} and form a practical stopping criterion.

The most popular bound proposed by Harrington~\cite{Harrington_TimeHarmonicElmagField} is based on the truncation of spherical wave series depending on the electrical size of an antenna. This estimate considers so-called \Quot{normal} gain, in which highly reactive components are not considered. This bound is solely determined by electrical size, which makes it easy to apply. Inevitably, such an estimate is not tight (except for the current density distribution on a spherical shell) and is overly conservative for high-gain applications. Still, knowing the number of available modes is helpful; therefore, the original work has been refined several times~\cite{Kildal+etal2017, Ziolkowski2017, Debard2023}, with all cases dealing with the maximization of antenna directivity constrained by additional parameters (bandwidth, ohmic losses, electrical size). The evaluation of the upper bound on directivity is extended towards arbitrarily-shapes obstacles in~\cite{Li2019} and \cite{PoloLopez2019} utilizing characteristic modes of perfectly conducting bodies and modal field expansions, respectively. The empiric truncation of the divergent series is, however, still required.

Current-density-based bounds were introduced recently~\cite{JelinekCapek_OptimalCurrentsOnArbitrarilyShapedSurfaces, GustafssonCapek_MaximumGainEffAreaAndDirectivity}, inspired by a similar approach for antenna arrays~\cite{UzsokySolymar_TheoryOfSuperDirectiveLinearArrays, HarringtonMautz_ControlOfRadarScatteringByReactiveLoading}. The current-based bounds are determined for an arbitrarily-shaped design region and a specified material distribution, providing a tight bound on antenna gain. The radiation modes were utilized~\cite{GustafssonCapek_MaximumGainEffAreaAndDirectivity} to interpret the composition of the optimal current solution, nevertheless, this composition is related to the optimal current only, and does not say much about a particular design region or its compatibility with excitation.

To overcome the above-mentioned issues, this paper shows how to decompose an optimal current generating upper bound on antenna gain into lossy characteristic modes~\cite{HarringtonMautzChang_CharacteristicModesForDielectricAndMagneticBodies}. The main result of this contribution is to show that the following formula holds
\begin{equation}
	\label{eq:GmaxVsCMA}
	\Gopt = \sum_n G_n,
\end{equation}
\ie{}, the upper bound on gain can be composed mode-by-mode from the characteristic gains. No antenna nor array constructed within the analyzed region can surpass this upper bound~\cite{Schab_SubstructureBounds}. Apart from the immediate applicability of~\eqref{eq:GmaxVsCMA}, the goal of this paper is to show that additional information can be gathered and used to assist with antenna design, \eg{}, the distribution of modal significance, optimal excitation coefficients, modal radiation efficiencies, or modal Q-factors. These modal parameters can also be used to truncate the series to restrict the upper bound further.

In accordance with the previous works, lossy conductors and dielectrics are considered in this work. The modal decomposition is utilized to diagonalize the dissipated power\footnote{In this work the word ``loss'' is used in the context of energy transformed into heat, while the word ``dissipation'' here covers both, the energy transformed into heat and radiated energy.} matrix in order to write the upper bound on antenna gain in the form of~\eqref{eq:GmaxVsCMA}. Since its advantageous algebraic properties, the modification of characteristic mode decomposition from~\cite{HarringtonMautzChang_CharacteristicModesForDielectricAndMagneticBodies} is utilized for lossy obstacles. As compared to characteristic modes for perfectly conducting bodies~\cite{HarringtonMautz_TheoryOfCharacteristicModesForConductingBodies}, where far fields are orthogonalized, the diagonalization of total dissipated (radiated plus ohmic lost) power is performed in this paper. The system matrix is still diagonalized, preserving the standard modal superposition formula when excitation is considered. This formulation is tractable, \eg{}, with an electric field integral equation implemented for surfaces or volumes or, in general, whenever the dissipated power can be written as a matrix operator. No separation of radiated power and power lost in heating is required.

The paper is organized as follows. The upper bound on antenna gain is provided in Section~\ref{sec:maxGain}. The dissipated power matrix from the denominator of the gain formula is decomposed in Section~\ref{sec:LCMA}, which makes it possible to write the upper bound on antenna gain in the form of~\eqref{eq:GmaxVsCMA} in Section~\ref{sec:sumRule}, \ie{}, to show that the upper bound is a sum of modal gains. The relation to the optical theorem is given in Section~\ref{sec:opticTheorem}. The theory is verified and its properties are demonstrated on an example of two parallel plates and end-fire arrays in Sections~\ref{sec:Ex1} and~\ref{sec:Ex2}, respectively. The additional aspects are treated afterwards: the upper bound for arbitrary polarization in Section~\ref{sec:highTotalGainX} and the constraint of self-resonance in Section~\ref{sec:selfResonance}. The paper is concluded in Section~\ref{sec:Conclu}.

\section{The Upper Bound on Antenna Gain}
\label{sec:maxGain}

Antenna gain is defined as~\cite[(2-46)]{Balanis_Wiley_2005}
\begin{equation}
	\label{eq:gain}
	G \left(\UV{e}, \UV{r}\right) = \dfrac{4\pi U \left(\UV{e}, \UV{r} \right)}{P} = \dfrac{2 \pi}{Z_0}\dfrac{\left| F \left(\UV{e}, \UV{r}\right) \right|^2 }{\Prad + \Plost},
\end{equation}
where $U \left(\UV{e}, \UV{r}\right) = \left| F \left(\UV{e}, \UV{r}\right) \right|^2 /(2 Z_0)$ is radiation intensity towards direction~$\UV{r}$ and of polarization~$\UV{e}$, $F \left(\UV{e}, \UV{r}\right)$ is corresponding electric far field, $P$ is the time-averaged dissipated power (total power accepted by the antenna), $\Prad$~is time-averaged radiated power, $\Plost$~is time-averaged lost power~(Joule heating), and $Z_0$ is a free-space impedance. The electric far field~$F \left(\UV{e}, \UV{r}\right)$ can also be written as
\begin{equation}
    \label{eq:Ffield}
    F \left(\UV{e}, \UV{r}\right) = \Fvec \left(\UV{e}, \UV{r}\right) \Ivec
\end{equation}
with~$\Fvec$ being a far-field matrix~\cite[(39)]{capek_etal_CMDusingScatteringDyadic} and $\Ivec = [I_p] \in \mathbb{C}^{N \times 1}$ being a column vector of expansion coefficients with the current density approximated with basis functions $\left\{\basisFcn_p \right\}$,~\cite{PetersonRayMittra_ComputationalMethodsForElectromagnetics} as
\begin{equation}
    \V{J}\left(\V{r}\right) \approx \sum_{p=1}^P I_p \basisFcn_p \left(\V{r}\right).
\end{equation}
 The antenna gain can be rewritten as
 \begin{equation}
	\label{eq:maxGainGEP}
	G \left(\UV{e}, \UV{r}\right) = \dfrac{4 \pi}{Z_0} \dfrac{\Ivec^\herm \Fvec^\herm \left(\UV{e}, \UV{r}\right) \Fvec \left(\UV{e}, \UV{r}\right) \Ivec}{\Ivec^\herm \left(\Rmat + \Lmat \right) \Ivec},
\end{equation}
where $\Rmat + \Lmat$ is a suitable model of total dissipated power. Within the electric field integral equation (EFIE) used in this paper, matrix~$\Rmat$ is the vacuum (radiation) part of the impedance matrix, and matrix~$\Lmat$ provides material losses~\cite{JelinekCapek_OptimalCurrentsOnArbitrarilyShapedSurfaces}. Within this paradigm, high-conducting surfaces are modeled with surface resistivity~$R_\T{s}$~\cite{SenoirVolakis_ApproximativeBoundaryConditionsInEM}, while dielectric materials are modeled using volume resistivity~$\rho$, leading to an element-wise definition of matrix~$\Lmat$ given as
\begin{equation}
    \label{eq:ohmiLossMatrix}
    L_{pq} = \int\limits_\varOmega \rho \left(\V{r}\right) \basisFcn_p^\ast \left( \V{r} \right) \cdot \basisFcn_q \left( \V{r} \right) \D{\varOmega}.
\end{equation}

The maximization of the antenna gain is derived in~\cite[(10-46)]{Harrington_FieldComputationByMoM} for terminal voltages and in~\cite{GustafssonCapek_MaximumGainEffAreaAndDirectivity} for ideal currents found in the presence of the lossy conductor, and reads
\begin{equation}
	\label{eq:GupFromIopt}
	\Gopt \left(\UV{e}, \UV{r}\right) = \dfrac{4 \pi}{\ZVAC} \Fvec \left(\UV{e}, \UV{r}\right) \left( \Rmat + \Lmat \right)^{-1} \Fvec^\herm \left(\UV{e}, \UV{r}\right).
\end{equation}
Optimal current~$\Iopt$ associated with the upper bound~$\Gopt$ is
\begin{equation}
	\label{eq:GainIopt}
	\Iopt = c \left(\Rmat + \Lmat\right)^{-1} \Fvec^\herm \left(\UV{e}, \UV{r}\right).
\end{equation}
with
\begin{equation}
    \label{eq:cConst}
	c = \dfrac{1}{\sqrt{\Fvec \left(\UV{e}, \UV{r}\right) \left(\Rmat + \Lmat\right)^{-1}\Fvec^\herm \left(\UV{e}, \UV{r}\right)}}.
\end{equation}
Formula~\eqref{eq:GupFromIopt} is the starting point in this paper to derive \eqref{eq:GmaxVsCMA}.

\section{Factorization of Dissipated Power Operator}
\label{sec:LCMA}

The factorization of matrix $\Rmat + \Lmat$ in \eqref{eq:GupFromIopt} makes it possible to express the upper bound as a sum of modal contributions. To this end, any decomposition involving the diagonalization of the dissipated power matrix can be involved, \eg{}, radiation modes were proposed in~\cite{GustafssonCapek_MaximumGainEffAreaAndDirectivity}. The decomposition proposed in \cite[(63)]{HarringtonMautzChang_CharacteristicModesForDielectricAndMagneticBodies} is utilized here for its link to the excitation. These \Quot{lossy characteristic modes} are defined\footnote{The authors are aware that more possible definitions of characteristic modes for lossy obstacles exist, \cf{}, \cite{HarringtonMautzChang_CharacteristicModesForDielectricAndMagneticBodies}. Here, we apply \eqref{eq:CMA} because of the advantageous algebraic properties of the formula to derive~\eqref{eq:GmaxVsCMA}, regardless of the naming conventions.} as
\begin{equation}
	\label{eq:CMA}
	\Xmat \Ivec_n = \lambda_n \left(\Rmat + \Lmat \right) \Ivec_n,
\end{equation}
where~$\Xmat \in \mathbb{C}^{N\times N}$ is the reactance matrix, \ie{}, the imaginary part of the system matrix within EFIE~\cite{JelinekCapek_OptimalCurrentsOnArbitrarilyShapedSurfaces}. Note that no separation of radiated power and power lost in heating is required since the entire matrix~$\Rmat + \Lmat$ is diagonalized. The normalization of current vectors~$\Ivec_n$ is performed so that
\begin{align}
	\Ivec_m^\herm \left(\Rmat + \Lmat \right) \Ivec_n &= \delta_{mn}, \label{eq:CMAnorm}\\
    \Ivec_m^\herm \Xmat \Ivec_n &= \delta_{mn} \lambda_n,
    \label{eq:CMAnormX}
\end{align} 
where~$^\herm$ is a Hermitian conjugate. The eigenvalues~$\lambda_n$ are real-valued, which is a consequence of solving a symmetric and real-valued generalized eigenvalue problem~\eqref{eq:CMA}.

The modes of~\eqref{eq:CMA} can be superposed as~\cite{HarringtonMautz_TheoryOfCharacteristicModesForConductingBodies}
\begin{equation}
	\label{eq:excitCoef}
	\Ivec = \sum_{n=1}^N \alpha_n \Ivec_n = \IVEC \V{\alpha} = \sum_n \dfrac{\Ivec_n^\herm \Vvec}{1 + \J \lambda_n} \Ivec_n,
\end{equation}
where~$\IVEC \in \mathbb{R}^{N\times N}$ is a matrix of modal currents ordered column by column, $\V{\alpha} \in \mathbb{C}^{N \times 1}$ is a vector of excitation coefficients, and where~$\Ivec \in \mathbb{C}^{N \times 1}$ is the \Quot{fed current} normally found via solution to the method-of-moments problem
\begin{equation}
    \label{eq:MoMeq}
	\Ivec = \Zmat^{-1} \Vvec,
\end{equation}
with~$\Vvec \in \mathbb{C}^{N\times 1}$ being the vector representing excitation (a delta gap, a plane wave, \dots). 

The key feature of the modes defined by~\eqref{eq:CMA}, which will be utilized here, is that
\begin{equation}
    \label{eq:Rinv}
    \left(\Rmat + \Lmat \right)^{-1} = \sum \limits_n \Ivec_n \Ivec_n^\herm = \IVEC \IVEC^\herm.
\end{equation}
The inversion~\eqref{eq:Rinv} exists since $\Lmat$ is a full-rank matrix for any real material, providing that the superposition~\eqref{eq:excitCoef} converges. The relation~\eqref{eq:Rinv} does not hold for classical characteristic modes~\cite{HarringtonMautz_TheoryOfCharacteristicModesForConductingBodies} due to rank-deficiency of matrix~$\Rmat$.

\section{Maximum Gain as a Sum of Lossy Characteristic Modes}
\label{sec:sumRule}

Considering formulas~\eqref{eq:GupFromIopt} and~\eqref{eq:Rinv}, the upper bound on antenna gain can also be written as
\begin{equation}
	\label{eq:GupFromCMAx}
	\Gopt \left(\UV{e}, \UV{r}\right) = \dfrac{4 \pi}{\ZVAC} \sum \limits_n \Ivec_n^\herm \Fvec^\herm \left(\UV{e}, \UV{r}\right) \Fvec \left(\UV{e}, \UV{r}\right) \Ivec_n  = \sum \limits_n G_n \left(\UV{e}, \UV{r}\right),
\end{equation}
where~$G_n$ represents characteristic gains
\begin{equation}
	\label{eq:gainCMA}
	G_n \left(\UV{e}, \UV{r}\right) = \dfrac{4 \pi}{\ZVAC} \dfrac{\Ivec_n^\herm \Fvec^\herm \left(\UV{e}, \UV{r}\right) \Fvec \left(\UV{e}, \UV{r}\right) \Ivec_n}{\Ivec_n^\herm \left(\Rmat + \Lmat \right) \Ivec_n}
\end{equation}
and where~\eqref{eq:CMAnorm} was applied. The formula~\eqref{eq:GupFromCMAx} is one of the main results of this paper and can be equivalently expressed as
\begin{equation}
	\label{eq:GupFromCharFields}
	\Gopt \left(\UV{e}, \UV{r}\right) = \dfrac{4 \pi}{\ZVAC} \sum \limits_n |F_n \left(\UV{e}, \UV{r}\right)|^2,
\end{equation}
where~$F_n \left(\UV{e}, \UV{r}\right)$ are characteristic far fields and where normalization~\eqref{eq:CMAnorm} is used. The highest possible gain is always produced by the sum of characteristic far fields squared and evaluated for the entire design region. This is a general statement, valid for all directions and frequencies, for a given arbitrarily shaped design region and material distribution. The value~\eqref{eq:GupFromCharFields} cannot be surpassed by any antenna, array, or current in a vacuum situated in a prescribed region, and, as such, it provides the antenna designer with important information at the beginning of the design process.

If demanded, the optimal current density representing the upper bound on antenna gain can also be decomposed into modes~\eqref{eq:CMA} as
\begin{equation}
	\label{eq:IoptVsIcm}
	\Iopt = \sum_n \beta_n \Ivec_n,
\end{equation}
where, by means of~\eqref{eq:GainIopt} and \eqref{eq:Rinv}, the expansion coefficients read
\begin{equation}
	\label{eq:betaCoef}
	\beta_n = c \left( \Fvec \left(\UV{e}, \UV{r}\right) \Ivec_n \right)^\herm = c F_n^* \left(\UV{e}, \UV{r}\right),
\end{equation}

The value of the maximum gain~$\Gopt$ is identical to the previous works~\cite{JelinekCapek_OptimalCurrentsOnArbitrarilyShapedSurfaces, GustafssonCapek_MaximumGainEffAreaAndDirectivity}, \ie{}, to~\eqref{eq:GupFromIopt}, however, formula~\eqref{eq:GupFromCMAx} opens new possibilities for classifying the number of modes required, how they couple to the used excitation, and how the bound~$\Gopt$ is reduced when some modes are not excited properly. Decomposition into characteristic modes also offers proof of the superiority of upper bound~$\Gopt$ to all antenna gains of realized designs corresponding to the same design region, see Appendix~\ref{sec:App0}. The inequality $G \left(\UV{e}, \UV{r}\right) \leq \Gopt \left(\UV{e}, \UV{r}\right)$ is connected to the fact that maximum antenna gain is realized by specific expansion coefficients~$\beta_n$ which are related to excitation vector~$\Vvec_\T{ub} = \Zmat \Iopt$ giving us an alternative way to get~\eqref{eq:betaCoef} from~\eqref{eq:excitCoef} by substituting~\eqref{eq:Rinv} and considering~\eqref{eq:CMAnorm}--\eqref{eq:CMAnormX}
\begin{equation}
\label{eq:alphaUb}
\Iopt = \sum_n \dfrac{\Ivec_n^\herm \Zmat \Iopt}{1 + \J \lambda_n} \Ivec_n = \sum_n c \left( \M{K} (\UV{e}, \UV{r}) \Ivec_n \right)^\herm  \Ivec_n = \sum_n \beta_n \Ivec_n.
\end{equation}

\section{Relationship to Forward Scattering and Optical Theorem}
\label{sec:opticTheorem}

The possibility to write the upper bound on antenna gain as a sum of lossy characteristic modes has a consequence in scattering theory. A balance between delivered power and dissipated power can be expressed in terms of the optical theorem~\cite{Kristensson_ScatteringBook}
\begin{equation}
    \label{eq:optTheorem}
    \Ivec^\herm \left( \Rmat + \Lmat \right) \Ivec = \RE \left\{ \Ivec^\herm \Vvec \right\} = -\dfrac{4\pi}{k \ZVAC} \IM \left\{ \Fvec \left(\UV{e}, \UV{r}\right) \Ivec \right\},
\end{equation}
where we utilize the relation between excitation coefficients~$\Vvec$ representing an impinging plane wave of unitary amplitude $\vert E_0 \vert = 1\,\T{Vm}^{-1}$, polarization~$\UV{e}$, and direction~$\UV{r}$, and far-field vector~$\Fvec \left(\UV{e}, \UV{r}\right)$ pointing towards the same direction~\cite[Eq. (41)]{capek_etal_CMDusingScatteringDyadic}, \ie{},
\begin{equation}
    \label{eq:ffVector}
    \Vvec = -\J \dfrac{4 \pi}{k \ZVAC} \Fvec^\herm (\UV{e}, \UV{r}).
\end{equation}

A question of maximum far-field intensity~$\vert F(\UV{e}, \UV{r})\vert^2$ subjected by optical theorem~\eqref{eq:optTheorem} is expressed as
\begin{equation}
\begin{aligned}
	& \mathrm{maximize} && \vert F(\UV{e}, \UV{r})\vert^2 \\
	& \mathrm{subject\,\,to} &&  \Ivec^\herm \left( \Rmat + \Lmat \right) \Ivec - \RE \left\{ \Ivec^\herm \Vvec \right\} = 0 \\
    & && \Vvec + \J \dfrac{4 \pi}{k \ZVAC} \Fvec^\herm (\UV{e}, \UV{r}) = 0
\end{aligned}
\label{eq:maxIntensityProb}
\end{equation}
and is solved in \cite[Eq. (55)]{2020_Gustafsson_NJP}. It can be shown that the optimal current solving~\eqref{eq:maxIntensityProb} is
\begin{equation}
    \Ivec = -\J \dfrac{4 \pi}{k \ZVAC} \left( \Rmat + \Lmat \right)^{-1} \Fvec^\herm \left(\UV{e}, \UV{r}\right)
    \label{eq:maxIntSol}
\end{equation}
which is the same current\footnote{The currents~\eqref{eq:maxIntSol} and \eqref{eq:GainIopt} differ only in amplitude, \ie{}, a constant. This constant is, however, irrelevant for the result of~\eqref{eq:gain}.} as for the upper bound on the antenna gain, \cf{} \eqref{eq:GainIopt}. In other words, the maximal forward scattering, when constrained by the optical theorem, can also be determined from the sum of characteristic gains~\eqref{eq:GupFromCMAx}. The scattering properties of an obstacle are fully characterized by a set of lossy characteristic modes, and their knowledge immediately gives an upper bound on forward scattering.

An interesting asymptotic relation can be found for electrically large obstacles ($ka \gg 1$ with $a$ being the smallest radius of the sphere fully circumscribing the antenna) by comparing characteristic mode expansion of the optimal current density \eqref{eq:IoptVsIcm} and the current density~\eqref{eq:excitCoef} excited by the plane wave in a forward scattering scenario, \ie{}, when~\eqref{eq:ffVector} holds. Substituting~\eqref{eq:ffVector} into~\eqref{eq:excitCoef} readily gives
\begin{equation}
    \label{eq:optCurLargeKA}
    \Ivec = -\J \dfrac{4\pi}{k \ZVAC} \sum_n \dfrac{1}{1 + \J \lambda_n} F_n^* \left(\UV{e}, \UV{r}\right) \Ivec_n.
\end{equation}
It is seen that for~$\vert \lambda_n \vert \longrightarrow 1$, $\forall n$, the expansion~\eqref{eq:optCurLargeKA} is equivalent to~\eqref{eq:IoptVsIcm}--\eqref{eq:betaCoef} (except for a constant), \ie{}, the current excited by the impinging plane wave is the same as the current maximizing forward scattering.

\section{Example: Maximum Gain of Two Parallel Plates}
\label{sec:Ex1}

\begin{figure}
\centering
\includegraphics[width=0.8\columnwidth]{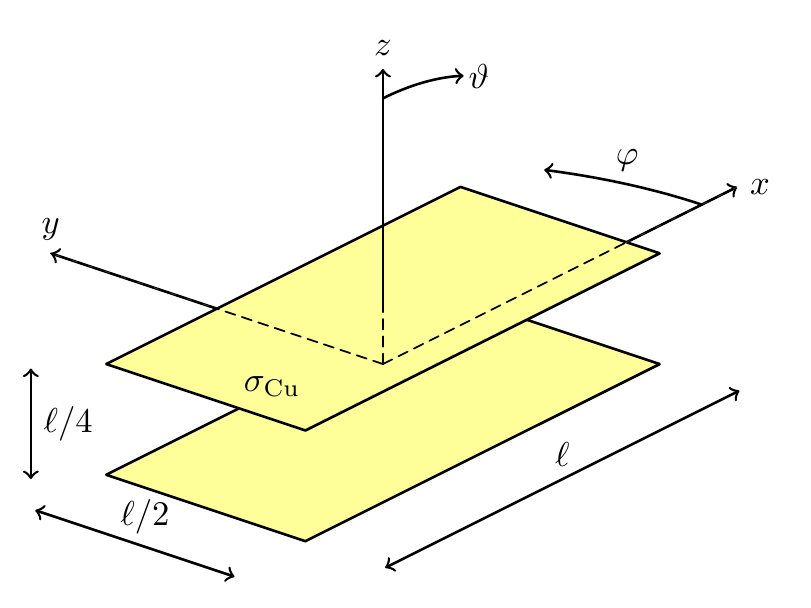}
\caption{The designing region studied in this paper: two parallel rectangular plates are centered in the coordinate system with edges being parallel to coordinate axes. The maximum gain is evaluated in the~$\varphi = 0$ plane. Polarization~$\UV{e}$ is defined using spherical angles~$\UV{\vartheta}$ and $\UV{\varphi}$.}
\label{fig1}
\end{figure}

To demonstrate the most salient properties of the proposed modal decomposition of the upper bound on antenna gain, a current supporting region in the form of two rectangular plates placed in parallel above each other is assumed, see Fig.~\ref{fig1}. The maximum gain is studied in the broadside direction, while end-fire radiation is treated in Section~\ref{sec:Ex2}. The size of each patch is~$\ell \times \ell/2$ and the separation distance is~$\ell/4$. The electrical size is~$k \ell = \pi$, where $k$~is a wave-number in free space. The chosen operating frequency is 750\,MHz with~$\ell = 0.2$\,m. Both plates are made of copper ($\sigma = 5.96\cdot 10^7\,\T{Sm}^{-1}$), and the ohmic losses are approximated with the thin-sheet model of equivalent surface resistance~$R_\T{s} = 0.007 \, \Ohm$. The model is discretized with 800~triangles and 1140~Rao-Wilton-Glisson basis functions~\cite{RaoWiltonGlisson_ElectromagneticScatteringBySurfacesOfArbitraryShape}. All matrix operators are evaluated in AToM~\cite{atom}. It is worth noting that despite the simplicity of the current supporting region, the upper bound discussed below applies to an arbitrary radiating structure fitting this design region, \ie{}, to a number of practical planar antenna designs.

\subsection{Upper Bound as a Sum of Modal Gains}

\begin{figure}
\centering
\includegraphics[width=\columnwidth]{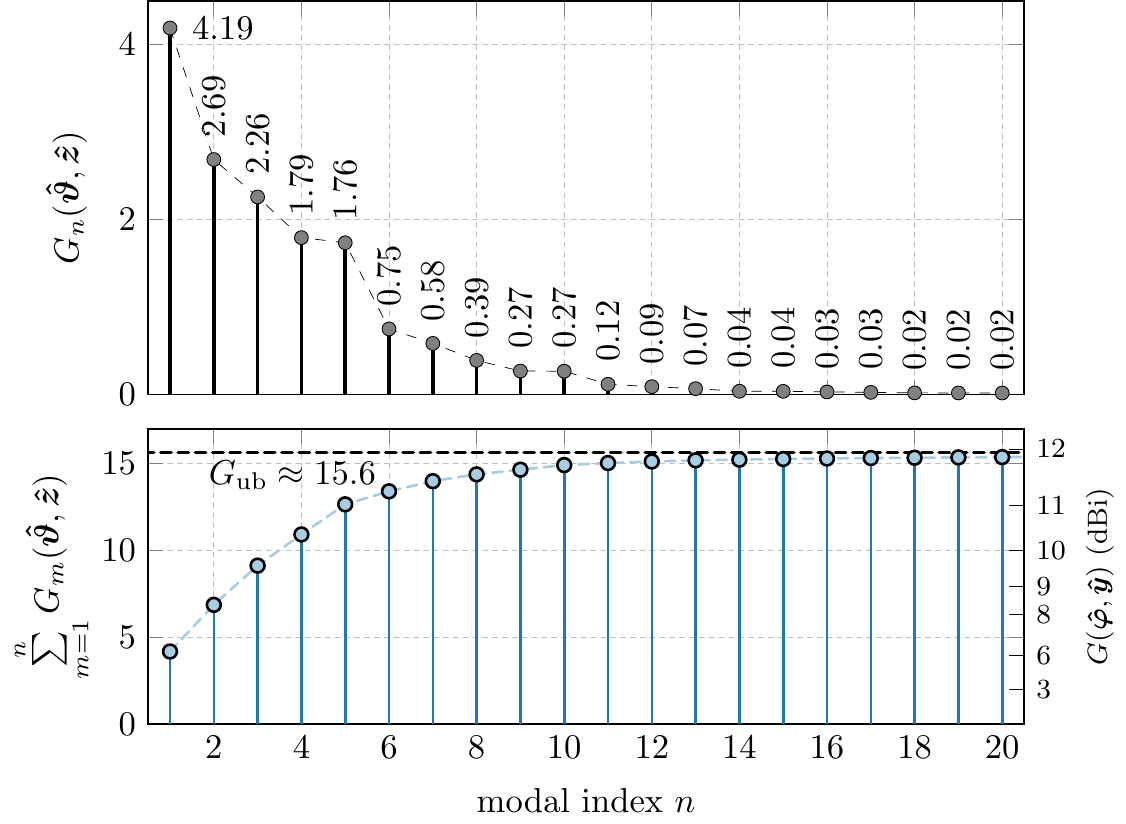}
\caption{Top: Modal antenna gains~$G_n$ for the setup depicted in Fig.~\ref{fig1}. Polarization~$\UV{e} = \UV{\vartheta}$, radiation direction~$\UV{r} = \UV{z}$, and electrical size~$k\ell = \pi$ are used. The modal gains are sorted from the highest to lowest in decreasing order. Bottom: The cumulative sum of modal gains. The sum of all modes gives the upper bound~$\Gopt$ (dashed line). Two y-axes are given for the bottom pane. The left y-axis depicts antenna gain in dimensionless units, while the right y-axis is in dBi. All numerical values within the panes are dimensionless.}
\label{fig2}
\end{figure}

The modal antenna gains~$G_n$ for the first 20~modes are depicted in the top pane of Fig.~\ref{fig2} and ordered (in descending fashion) according to gain. Only the first ten modes significantly contribute to the upper bound; their summation gives~$\approx 95.4\%$ of the bound~$\Gopt$ and all remaining modes each contribute less than $1\%$ to the bound. This is graphically depicted by the cumulative sum of modal antenna gains in Fig.~\ref{fig2}, bottom pane. Taking the sum of all modes yields the upper bound~$\Gopt$, \cf{} \eqref{eq:GupFromCMAx}.

\subsection{Performance in Other Modal Metrics}

\begin{figure}
\centering
\includegraphics[width=\columnwidth]{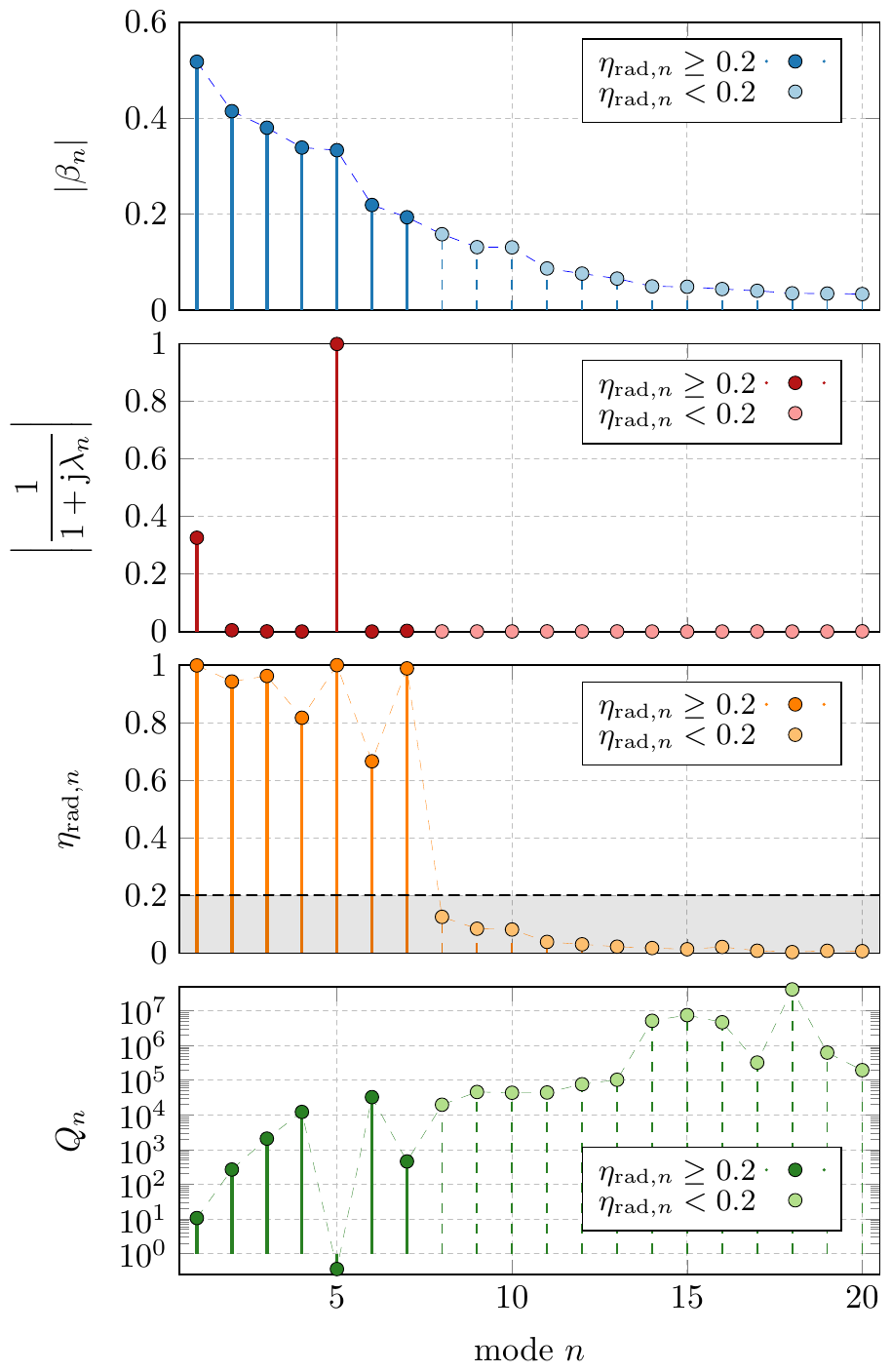}
\caption{Excitation coefficient~$\beta_n$, modal significance~$|1/(1+\J\lambda_n)|$, characteristic radiation efficiency~$\eta_{\T{rad},n}$, and characteristic radiation Q-factors~$Q_n$ for modal gains from Fig.~\ref{fig2}. The ordering is the same. The modes are further classified as normal ($\eta_{\T{rad},n} > 0.2$) and super-directive ($\eta_{\T{rad},n} \leq 0.2$). The distinction is made by markers of darker and lighter colors for all the panes.}
\label{fig3}
\end{figure}

The upper bound is not attainable in practice. This is because all modes cannot be excited together, each with its specific excitation coeficient~$\beta_n$, \cf{}, \eqref{eq:betaCoef}. To this point, the first 20~modes are analyzed in Fig.~\ref{fig3} with the same ordering as in Fig.~\ref{fig2}. In addition, the modes are separated into two groups, depending on their modal radiation efficiency
\begin{equation}
    \label{eq:CMAmodalEff}
    \eta_{\T{rad},n} = \dfrac{P_{\T{rad},nn}}{P_{\T{rad},nn} + P_{\T{lost},nn}} = \dfrac{\Ivec_n^\herm \Rmat \Ivec_n}{\Ivec_n^\herm \left(\Rmat + \Lmat \right) \Ivec_n} = \Ivec_n^\herm \Rmat \Ivec_n,
\end{equation}
where the normalization~\eqref{eq:CMAnorm} was considered in the denominator of the RHS of~\eqref{eq:CMAmodalEff}. The modes with modal efficiency lower than~20\% are called super-directive modes and are highlighted by the light-colored marks. The modes with modal efficiency~$\eta_{\T{rad},n}$ higher than~20\% are called normal modes and are highlighted by the dark-colored marks. It is seen that only the first seven modes have modal efficiency considerably higher than~20\%. Furthermore, only two modes (the 1st and 5th modes) have high modal significance at this electrical size ($k\ell = \pi$). The top pane shows the absolute value of the optimal excitation coefficient~$\vert \beta_n \vert$. The study in Fig.~\ref{fig3} reveals that two parallel plates have high potential in antenna gain, however, it is impossible in practice to achieve it. Realistically, a gain of around $6$ should be expected (summation of modal gains of the 1st and the 5th mode). Changing the topology of the design region might cause some modes to improve their excitation coefficients and/or other modal metrics but lowers the upper bound, which always decreases with any perturbation of the design domain. As a final remark to Fig.~\ref{fig3}, the modal Q factors, evaluated as~\cite{CapekHazdraEichler_AMethodForTheEvaluationOfRadiationQBasedOnModalApproach}
\begin{equation}
    Q_n = \dfrac{2 \omega \max \left\{W_{\T{m},n}, W_{\T{e},n}\right\}}{\Prad},
\end{equation}
where $W_{\T{m},n}$ and $W_{\T{e},n}$ are cycle mean magnetic and electric stored energies \cite{Vandenbosch_ReactiveEnergiesImpedanceAndQFactorOfRadiatingStructures}, respectively, are shown in the bottom pane. Only two modes (the 1st and 5th modes) exhibit a small Q-factor. This indicates that adding up more modes (boosting the gain) narrows the operational bandwidth of an antenna. This observation is well aligned with Harrington's conclusions~\cite{Harrington_TimeHarmonicElmagField} and will be commented on in detail at the end of this section.

\subsection{Optimal and Modal Currents}

\begin{figure*}
\centering
\includegraphics[width=0.4\textwidth]{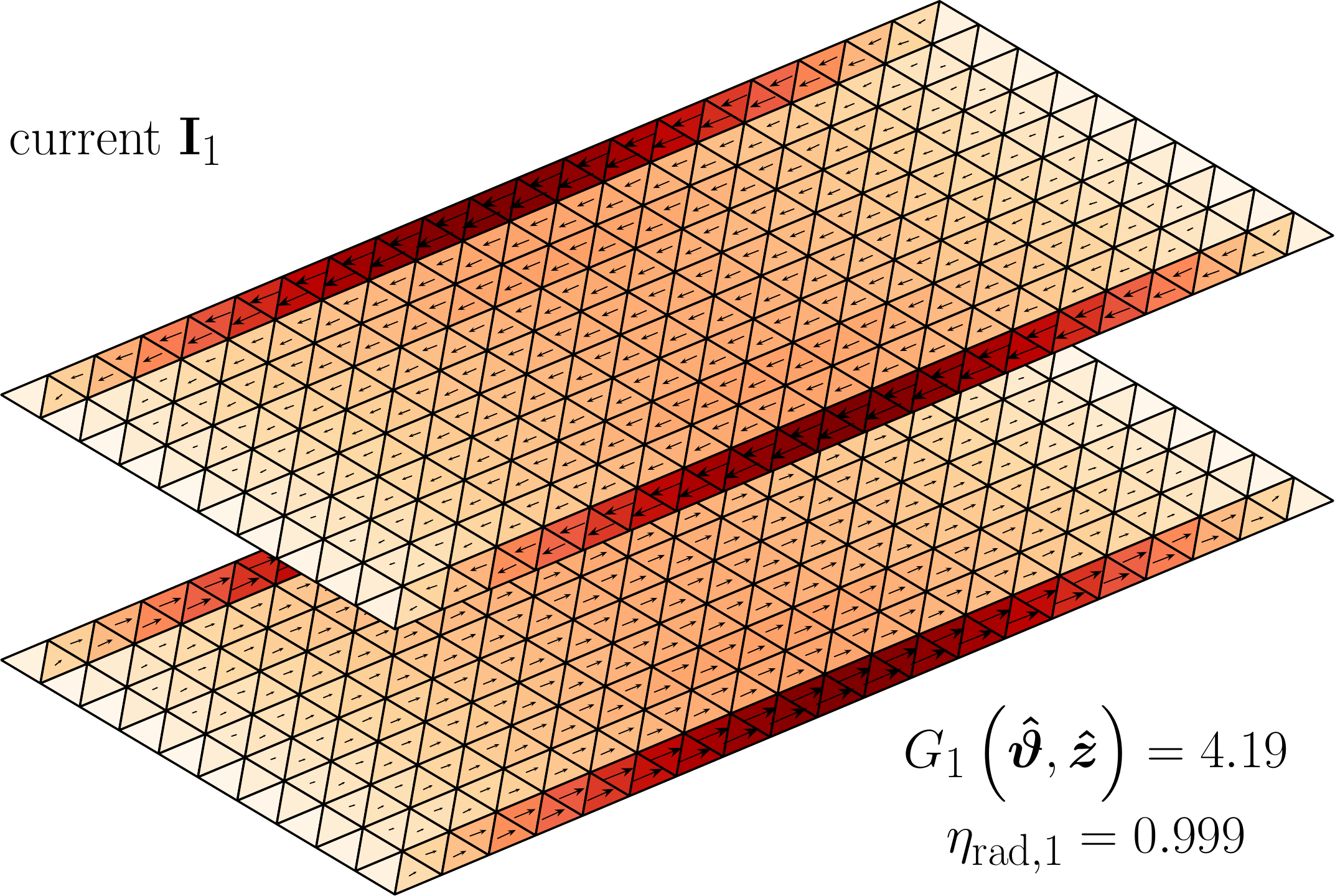}
\hspace{0.75cm}
\includegraphics[width=0.4\textwidth]{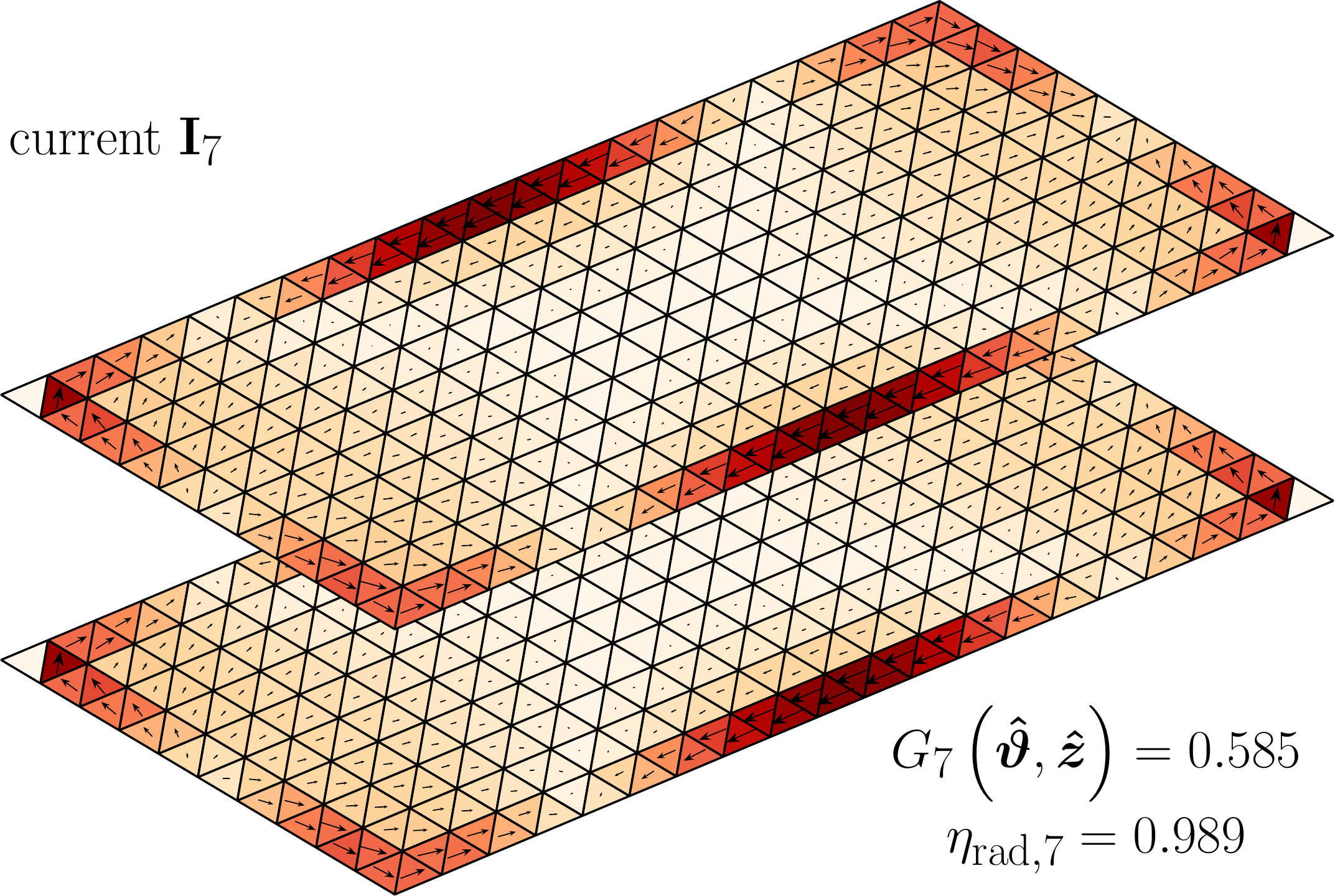}

\vspace{0.75cm}
\includegraphics[width=0.4\textwidth]{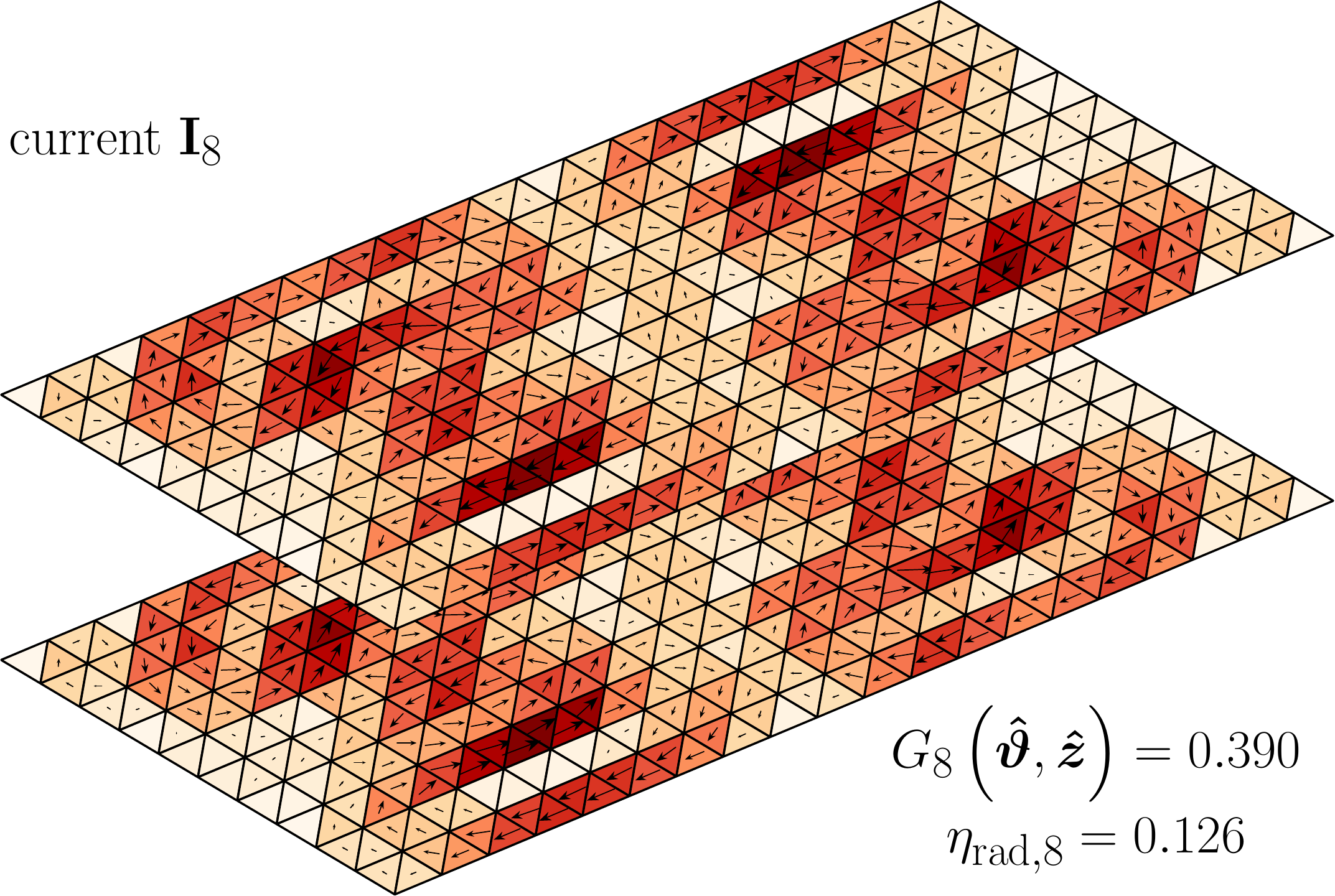}
\hspace{0.75cm}
\includegraphics[width=0.4\textwidth]{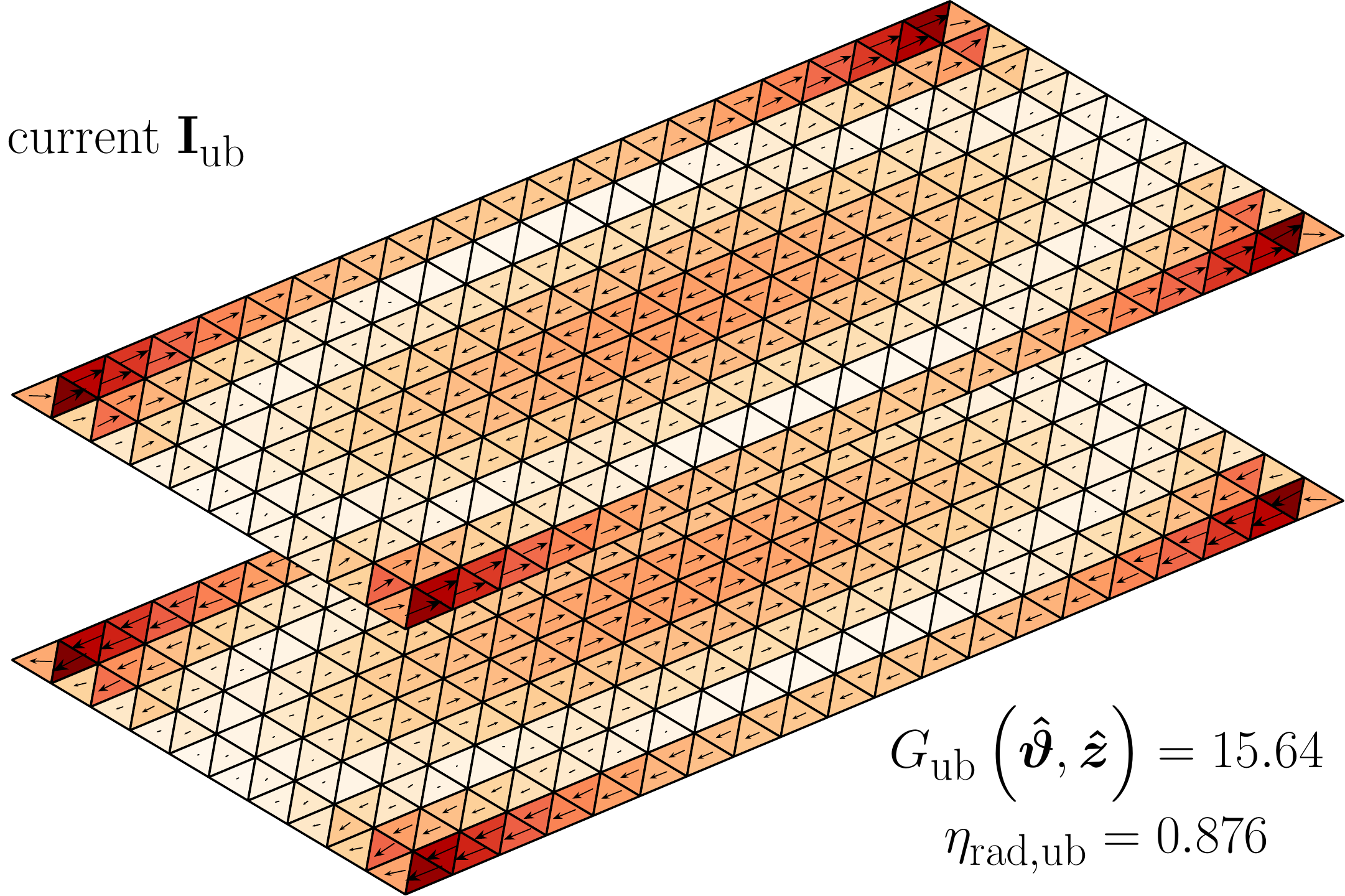}
\caption{Surface current density for three selected lossy characteristic modes of a structure from Fig.~\ref{fig1} and the optimal current density~$\Iopt$ (bottom-right corner). Current~$\Ivec_1$ has the highest modal antenna gain. Current~$\Ivec_7$ is the last \Quot{normal} mode (decent modal gain but high modal radiation efficiency), and current~$\Ivec_8$ is the first \Quot{super-directive} characteristic mode (decent modal gain but low modal efficiency).}
\label{fig4}
\end{figure*}

Characteristic modes are often used for their insight into radiation mechanisms. The intuition of antenna designers often stems from an inspection of modal currents. Several modal currents are shown in Fig.~\ref{fig4} and demonstrate their different properties. The dominant mode with the highest modal gain is shown in the top-left corner. The last normal mode (with $\eta_{\T{rad},n} \geq 0.2$) is the 7th mode, \cf{} Fig.~\ref{fig3} bottom pane, and is shown in the top-right corner. It can be seen that both the 1st and the 7th modes have smooth current distributions, therefore, they have high modal efficiency. The first super-directive mode, the 8th mode, is shown in the bottom-left corner. As expected, the current density is highly irregular, causing high ohmic losses. Finally, the optimal current density~$\Iopt$ is depicted in the bottom-right corner and resembles an array with dipole radiators in the corners.

\subsection{Angular Dependence of the Upper Bound}

\begin{figure}
\centering
\includegraphics[width=\columnwidth]{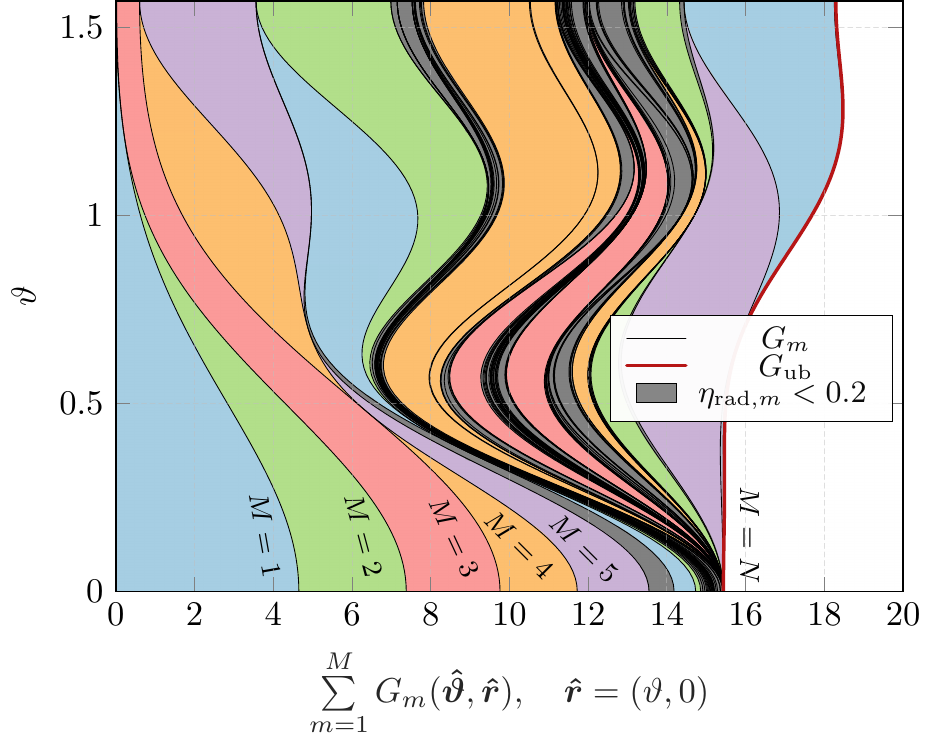}
\caption{The composition of modal gains ($x$-axis) depending on the angle of radiation ($y$-axis, given in $\vartheta$). Polarization $\UV{e} = \UV{\vartheta}$ is used. When all modes are summed up, $M= N$, the upper bound on antenna gain is realized. Summing fewer modes gives $G < \Gopt$. The modes are ordered as in Fig.~\ref{fig2}, \ie{}, according to modal gains for $\UV{r} = \UV{\vartheta}$ direction.}
\label{fig5}
\end{figure}

Different characteristic modes contribute to the upper bound on gain differently depending on direction and polarization. Figure~\ref{fig5} demonstrates this fact exactly by studying the dependence on elevation angle~$\vartheta$ for a given polarization~$\V{\vartheta}$ and azimuth $\varphi = 0$. Two sets of modes are, once again, distinguished, based on their modal efficiency -- normal modes are represented by colors, and super-directive modes are gray. The modes are sorted from the highest to the lowest modal gain for radiation direction~$\vartheta = 0$ and the graph shows antenna gain when $M$ modes are summed. It is seen, for example, that the mode of the highest modal gain ($M = 1$) for $\vartheta = 0$ does not contribute to the bound for~$\vartheta = \pi/2$. When all modes are summed, the upper gain is found (see the red thin line for $M = N$).

\begin{figure}
\centering
\includegraphics[width=\columnwidth]{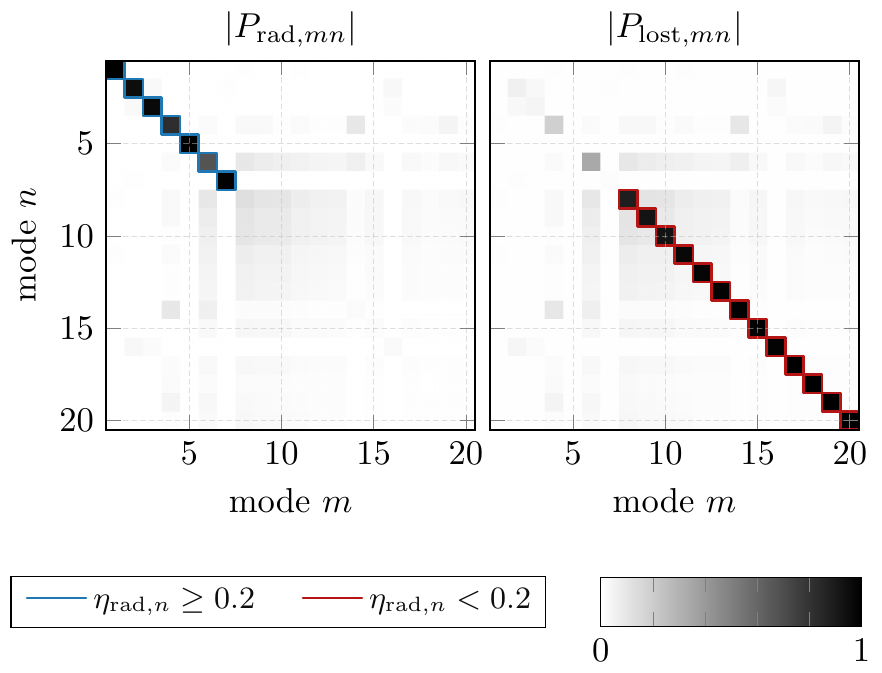}
\caption{Power terms~$P_{\T{rad},mn} = \Ivec_m^\herm \Rmat \Ivec_n / 2$ (left) and~$P_{\T{lost},mn} = \Ivec_m^\herm \Lmat \Ivec_n / 2$ (right). Absolute values are shown because of possible negative off-terms. The sum~$P_{\T{rad},mn} + P_{\T{lost},mn}$ is normalized according to~\eqref{eq:CMAnorm}, \ie{}, is one for $m = n$ and zero otherwise. While the separated terms are not orthogonalized by lossy characteristic modes, the correlation between different modes, $m \neq n$, is low for good conductors and low-order modes. It is seen that the first seven modes predominantly radiate, while the remaining modes cause high ohmic losses.}
\label{fig6}
\end{figure}

The modal radiated and lost power, including mutual terms~$\Ivec_m^\herm \Rmat \Ivec_n/2$, $m \neq n$, are depicted in Fig.~\ref{fig6}. The radiated power dominates for normal modes, while the lost power dominates for super-directive modes. It is also seen that neither radiated nor lost power is orthogonal in the case of lossy characteristic modes, \cf{} \eqref{eq:CMAnorm}. The correlation between different modes is, however, low for modes strongly contributing to the upper bound on gain.

\subsection{Upper Bound as a Function of Material Conductivity}

\begin{figure}
\centering
\includegraphics[width=\columnwidth]{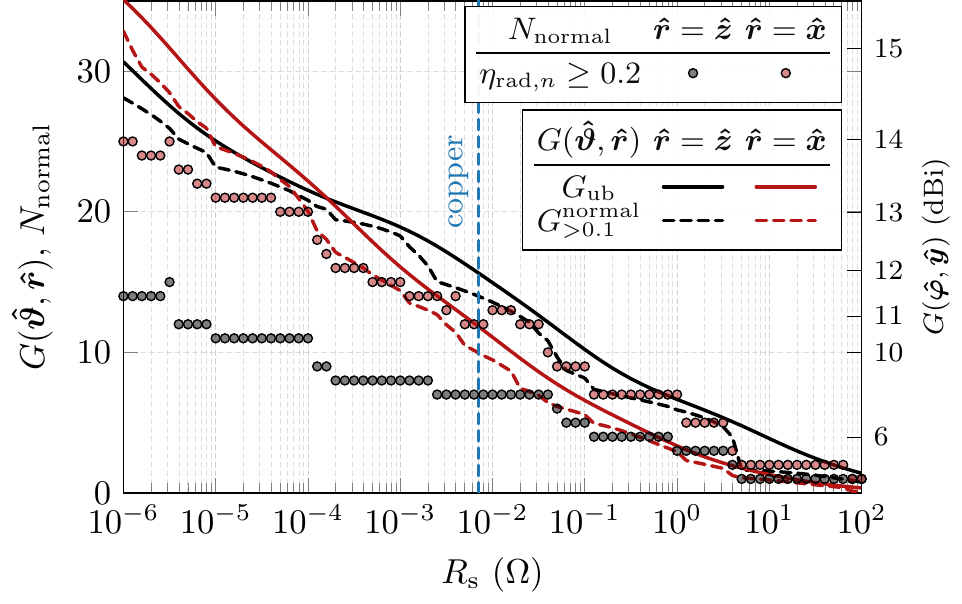}
\caption{Maximal antenna gain depending on the resistivity~$R_\T{s}$. Two directions of radiation are optimized and analyzed; black traces and marks are for~$\UV{r} = \UV{z}$, and the red traces and marks are for~$\UV{r} = \UV{x}$. Polarization $\UV{e} = \UV{\vartheta}$ is used. The solid lines show the upper bound on antenna gain. The circle marks show the number of normal modes~$N_\T{normal}$ participating in the upper bound with a value of modal gain ~$G_n > 0.1$. Summing modal gains of only these modes gives~$G_{>0.1}^\T{normal}$, depicted by the dashed lines.}
\label{fig7}
\end{figure}

The upper bound significantly depends on the properties of the materials used, see Fig.~\ref{fig7} showing the upper bound on antenna gain for two radiation directions, $\UV{r} = \UV{z}$ and $\UV{r} = \UV{x}$, and polarization $\UV{e} = \UV{\vartheta}$. A large range of surface resistivities~$R_\T{s}$ is considered, with copper ($R_\T{s} = 0.007\,\Ohm$) highlighted by the dashed vertical line. The upper bound~$\Gopt$ shown by the solid lines increases with decreasing resistivity and diverges for zero resistivity~$R_\T{s}$, as anticipated from the unbounded nature of the upper bound on directivity~\cite{Bloch_PIEE1953}.

The number of participating normal modes~$N_\T{normal}$ is represented by the circle marks. Only modes contributing at least 0.1 to the upper bound are considered. When only these modes are summed up as $G_{>0.1}^\T{normal}$, the dashed lines are obtained, indicating that this number of modes is sufficient to represent the upper bound in good precision. Notice a disproportion between the number of modes for $\UV{r} = \UV{z}$ (broadside) and $\UV{r} = \UV{x}$ (end-fire) directions, even though the upper bounds are similar. 

\subsection{Comparisons With Other Works}

\begin{figure}
\centering
\includegraphics[width=\columnwidth]{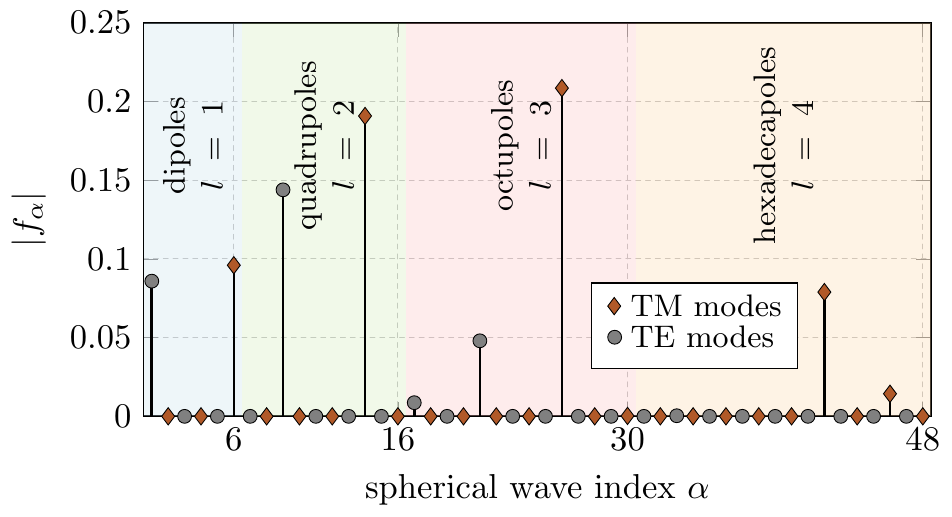}
\caption{Spherical wave expansion of the optimal current. The power radiated by each spherical wave component is $|f_\alpha|^2/2$, where $\alpha$ is a spherical wave multi-index~\cite{Hansen_SphericalNearFieldAntennaMeasurements, TayliEtAl_AccurateAndEfficientEvaluationofCMs}. The components are classified according to their order~$l$ and type (TM/TE).}
\label{fig8}
\end{figure}

The upper bound on antenna gain along the $z$-axis reaches~$\Gopt \approx 15.6$ (11.9\,dBi). This value is significantly higher than the prediction given by Harrington's estimate~\cite{Harrington_OnTheGainAndBWofDirectAntennas, Harrington_TimeHarmonicElmagField}
\begin{equation}
    \label{eq:HarrEst}
    G_\T{normal} = ka^2 + 2 ka,
\end{equation}
which yields~$G_\T{normal} \approx 6.84$ (8.35\,dBi). The estimate~\eqref{eq:HarrEst} is based upon the assumption that only spherical modes up to the order $\Lmax = 2$ (dipoles and quadrupoles) are used, since $\Lmax = \lceil ka \rceil$. Nevertheless, the spherical mode expansion of optimal current~$\Iopt$ gives a distribution shown in Fig.~\ref{fig8} where a combination of TM/TE doublets is used up to $\Lmax = 4$. Consequently, the remarkable difference between the upper bound predicted in this work and by an estimate given by Harrington, or similarly by Kildal \textit{et al.}~\cite{Kildal+etal2017}, is caused by highly reactive currents. As shown in the analysis above, see Fig.~\ref{fig3}, they have either high Q-factor or extremely low radiation efficiency, and they are usually not utilized in practice. As such, the bound presented here\footnote{The same conclusion holds for~\cite{GustafssonCapek_MaximumGainEffAreaAndDirectivity}, giving the same upper bound $\Gopt$, however, employing another computational approach without the possibility to sum up modal gains directly and to utilize additional information provided by modal analysis.} is the true bound, although it may look overly optimistic in the majority of cases, but is always higher than the actual performance of an antenna. This is not the case with Harrington's estimate, which can be surpassed by skilled antenna designers~\cite{Yaghjian_etal_RS2008, Shi2022} or automated inverse design procedures~\cite{Li2022}.

\begin{figure}
\centering
\includegraphics[width=\columnwidth]{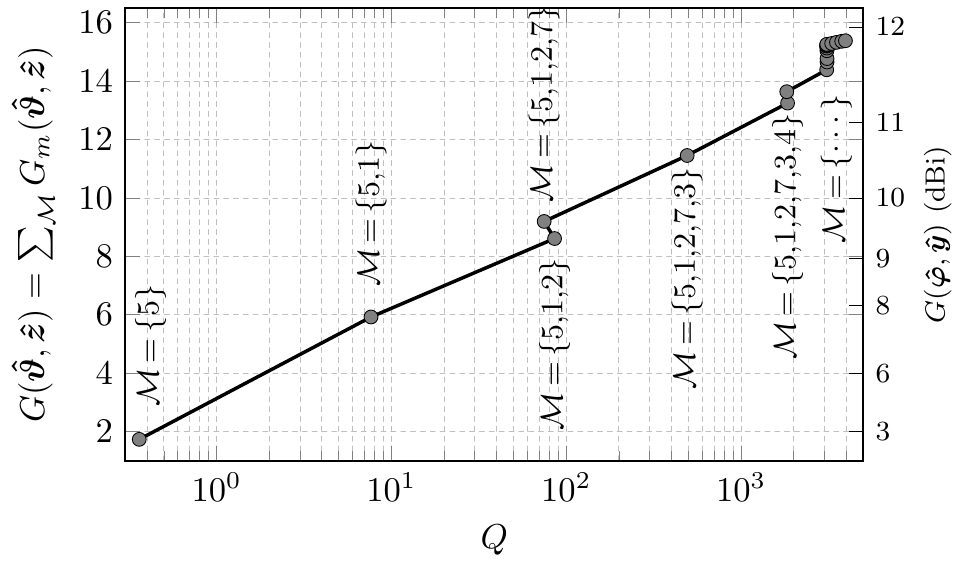}
\caption{Antenna gain~$G$ evaluated as a sum of characteristic modal gains with their optimal coefficients~$\V{\beta}$ to meet the upper bound~$\Gopt$. Only those modes indicated by set~$\OP{M}$ are summed. They are summed cumulatively depending on their modal Q-factors, starting from the mode with the lowest modal $Q_n$ (the 5th mode, see Fig.~\ref{fig3}, bottom pane).}
\label{fig9}
\end{figure}

The difference between the estimate~\eqref{eq:HarrEst} and the upper bound on gain~\eqref{eq:GupFromCMAx} can also be reconsidered, taking into account the Q-factor (which is approximately inversely proportional to fractional bandwidth~\cite{YaghjianBest_ImpedanceBandwidthAndQOfAntennas}). The first 20~characteristic modes, as shown in Fig.~\ref{fig2}, are summed with coefficients~$\V{\beta}$ from Fig.~\eqref{fig3}. The gain~$G$ is expressed as a cumulative sum of modal gains, adding up modes one by one and ordered according to their modal Q-factors, \cf{} Fig.~\ref{fig3}. Starting with the low-Q modes, the first two solutions with the lowest~$Q_n$ (and gain $G$) use the 5th mode and summation of the 5th mode and the 1st mode. The content of the sum is indicated by a set $\OP{M} = \left\{ m \right\}$. Figure~\ref{fig9} demonstrates the cost of high gain~$G$ in terms of Q-factor. Mixtures~$\OP{M}$ are not optimized for low~$Q$, since mixing coefficients~$\V{\beta}$ are taken from~\eqref{eq:betaCoef}. Therefore, high Q-factors might occur for high gain. This confirms (regardless of the shape of the design region) that antenna gain~$G$ higher than normal gain~\eqref{eq:HarrEst} (super-gain) is redeemed by a narrow bandwidth~\cite{Harrington_TimeHarmonicElmagField}.

The distribution of modal gains~$G_n$ to reach the upper bound~\eqref{eq:GupFromCMAx} is unique for each given shape of the design region, material, and frequency. The decomposition in lossy characteristic modes might indicate the principal degrees of freedom to use when antenna (inverse) design is applied. When the topology of an antenna is restricted to the subregion of an analyzed object, \eg{}, by cutting slots, the upper bound can only decrease, but the distribution of modal gains or other modal parameters (excitation coefficients, radiation efficiency, Q-factor) may improve. This is, however, not expected when the overall electrical size $ka$ is reduced.

\section{Performance of End-Fire Arrays}
\label{sec:Ex2}

The design of high-directivity end-fire arrays is extensively covered in the literature~\cite{Collin_Zucker_AntennaTheoryPart1, Clemente2015, Hazdra_etal_OnEndFireSuperDirectivityOfArrays, Yaghjian_etal_RS2008, Haskou2015, AltshulerODonnellYaghjianBest_AMonopoleSuperdirectiveArray}. Most of the contributions, however, attempt to optimize directivity, taking into account the influence of ohmic losses \textit{ex-post}, \cite{Debard2023}. A detailed treatment of the topic, including optimal excitation and upper limits, is given in~\cite[Ch.~5]{Collin_Zucker_AntennaTheoryPart1} regarding uniform arrays. We approach the problem of two- and three-element arrays in this section, focusing on the insight provided by the modal method, which considers particular shapes and ohmic losses. An optimal number of radiators, feeding placements, and proper excitation are studied as well.

\subsection{Upper Bound on the Design Region}

\begin{figure}
\centering
\includegraphics[width=\columnwidth]{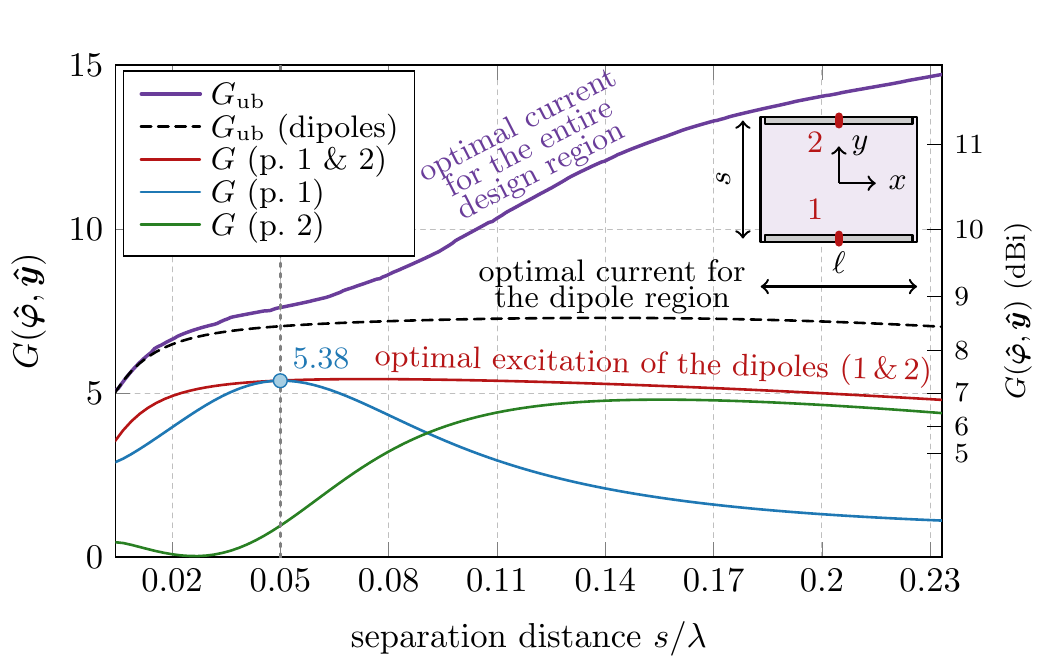}
\caption{The upper bound on antenna gain~$\Gopt$ of the design region of dimensions~$\ell \times (s + w)$, where~$\ell = \lambda/2 = 1$\,m and~$w = \ell/60$, is shown in purple. Other curves correspond to two dipoles of width~$w$, separation distance~$s$, and length~$0.945\ell$, which are placed at the opposite sides of this design region. The black dashed line shows the performance of the optimal current maximizing the gain when only the dipole region is used. When both dipoles are fed optimally~\cite{Capeketal_OptimalityOfTARCAndRealizedGainForMultiPortAntennas} in their centers, the red curve is obtained. The blue and green curves show scenarios where the 1st or the 2nd dipole is separately fed. The arrangements and the numbering of the dipoles are shown in the inset. The current supporting regions are made of copper.}
\label{fig10}
\end{figure}

In what follows, let us consider planar structures, lying in the $x$--$y$ plane, and radiating towards the direction of~$\UV{y}$ ($\vartheta = \pi/2$, $\varphi = \pi/2$) with field polarization~$\UV{e} = \UV{\varphi}$. As with the previous example, the material is copper.

To start, assume a rectangular region of width~$\ell = \lambda/2$ and variable height~$s+w$, where~$w=\ell/60$. The fundamental bound on the upper gain, $\Gopt (\UV{\varphi}, \UV{y})$, is investigated in Fig.~\ref{fig10}, see the purple curve. The maximal allowed height is~$\ell/2$ which gives the upper bound on gain~$\Gopt \approx 14.7$ (11.7\,dBi). No design realized within the~$\ell \times (s + w)$ region could outperform the corresponding limit value.

\subsection{End-Fire Array of Two Dipoles}
\label{eq:2dipPerf}

To be closer to antenna practice, assume that within the $\ell \times (s+w)$ design region, only two strip dipoles of length~$0.945 \ell$ (self-resonant length) and width $w = \ell/60$ are placed, and these are separated by center-to-center distance~$s$, see the inset in Fig.~\ref{fig10}. The dipoles have discrete delta gaps in their centers and are numbered~$1$ and~$2$ for the back and front, respectively. Several scenarios are considered.

The dashed black line in Fig.~\ref{fig10} shows the upper bound on antenna gain when only the conductive regions spanned by the dipoles are used to find the optimal currents. The red, blue, and green lines, respectively, show the performance when both dipoles are fed optimally; only dipole~$1$ is fed, and only dipole~$2$ is fed separately. It is seen that there is a specific distance, $s/\lambda \approx 0.05$, highlighted by the vertical dashed line, where only one feeder is enough to reach relatively high-gain performance~$G \approx 5.38$ ($7.31$\,dBi). 

\subsection{Modal Analysis of Two-Dipole Arrangement}
\label{eq:2dipmodes}

\begin{figure}
\centering
\includegraphics[width=\columnwidth]{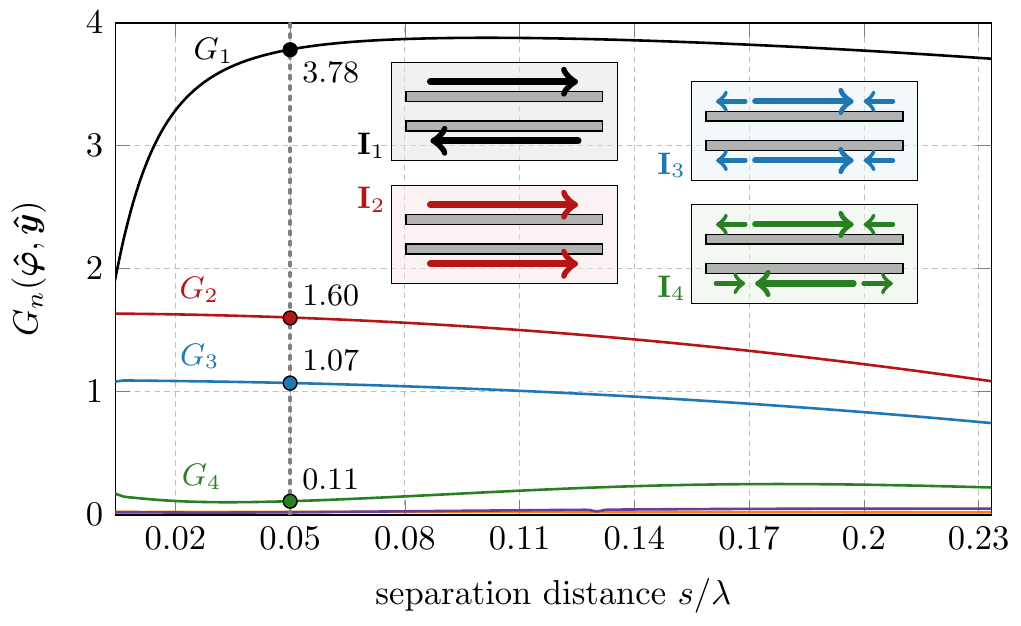}
\caption{Modal gains~$G_n$ of two dipoles for the same setup as in Fig.~\ref{fig10}. Only four modes have a non-negligible contribution to the upper bound on gain~$\Gopt$ shown in Fig.~\ref{fig10} as the black dashed line. The corresponding modal currents $\Ivec_1$--$\Ivec_4$ are depicted as the insets.}
\label{fig11}
\end{figure}

\begin{figure}
\centering
\includegraphics[width=\columnwidth]{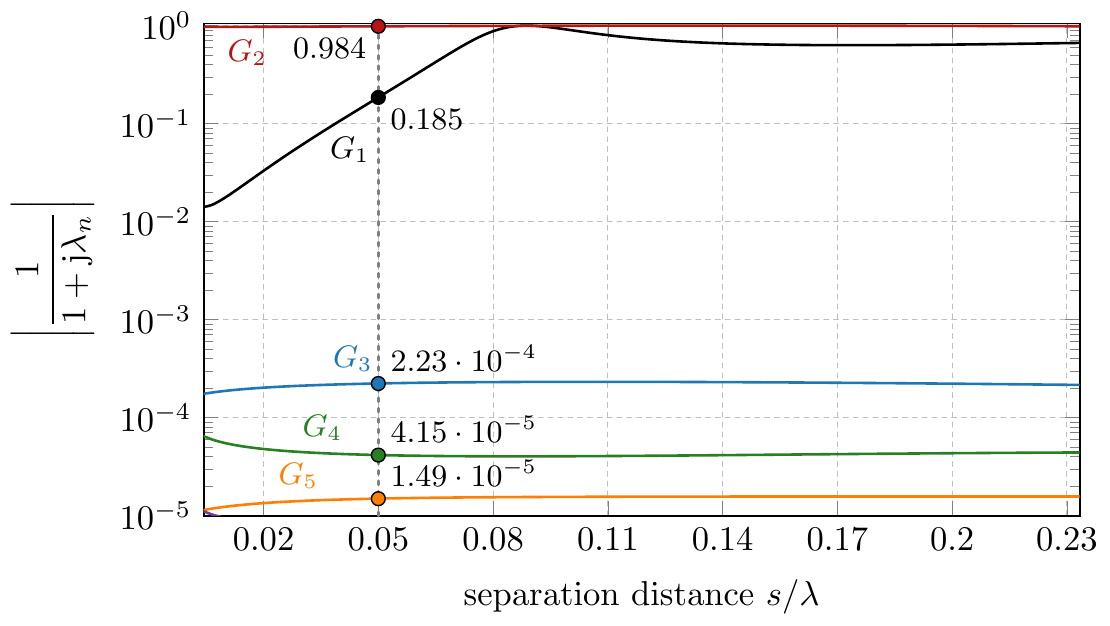}
\caption{Modal significance for the modes depicted in Fig.~\ref{fig11}. The same ordering and color-coding of the curves are used. Only two modes (with modal gains~$G_1$ and~$G_2$) have non-negligible modal significance, \ie{}, the ability to be excited.}
\label{fig12}
\end{figure}

The modal gains are depicted in Fig.~\ref{fig11}. Their summation gives the upper bound for the optimal current in the dipole region, depicted by the black dashed curve in Fig.~\ref{fig10}. There are only four modal gains in Fig.~\ref{fig11}, $G_1$--$G_4$, which contribute to the upper bound by at least~$1\,\%$. The associated current densities are sketched in the insets of Fig.~\ref{fig11}. Considering the separation distance~$s/\lambda = 0.05$ again, it can be seen that the gain realized by the dipoles with feeding placed at position 1 ($G \approx 5.38$) is exactly achieved when modes 1 and 2 ($G_1 \approx 3.78$, $G_2 \approx 1.60$) are perfectly excited. This fact is further supported by the study of modal significances in Fig.~\ref{fig12}, where it is seen that only modes~1 and~2 have high modal significance evaluated as~$\vert 1/(1 + \J\lambda_n) \vert$.

\begin{figure}
\centering
\includegraphics[width=\columnwidth]{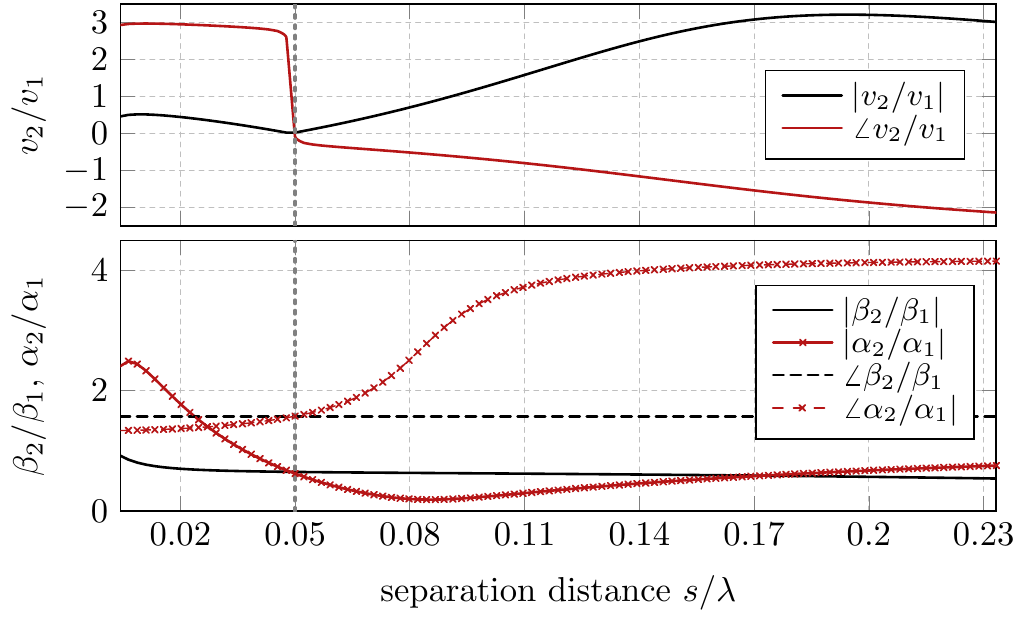}
\caption{The optimal voltage impressed to the voltage delta gaps placed in the middle of the 1st and 2nd dipoles, see the inset in Fig.~\ref{fig10}. The delta gap of the 2nd dipole can be shortened for separation distance~$s = 0.05 \lambda$, see the top pane. In this case, just a single delta gap connected to the 1st dipole is enough for the optimal excitation of the first two modes, see the bottom pane. The optimal excitation coefficients~$\beta_1$ and~$\beta_2$ are strictly followed with the excitation of the 1st dipole as shown by the coefficients~$\alpha_1$ and~$\alpha_2$, leading to gain~$G = G_1 + G_2 \approx 5.38$ (7.31\,dBi), see Fig.~\ref{fig10} at $s = 0.05 \lambda$.}
\label{fig13}
\end{figure}

Knowing that only two modes can be excited and that these modal gains give $G = 5.38$, we return to the optimal excitation of the dipoles. For sufficient freedom, both ports are fed optimally~\cite{Capeketal_OptimalityOfTARCAndRealizedGainForMultiPortAntennas}. The top pane of Fig.~\ref{fig13} reveals that $v_2 = 0$\,V at $s/\lambda = 0.05$, allowing the removal of port number~2. Leveraging the knowledge of optimal modal excitation coefficients~$\V{\beta}$, it is seen from the bottom pane that they are identical to the excitation coefficients~$\V{\alpha}$ of modes~$1$ and~$2$, \ie{}, $\beta_1 = \alpha_1$ and $\beta_2 = \alpha_2$.

The study above shows and explains why two dipoles of length $\ell \approx 0.473\lambda$ and separation distance~$0.05\lambda$ are capable of reaching a gain of $G \approx 5.38$ with only the back dipole being fed. The reason is that the dipole and quadruple modes are ideally excited. A question arises as to whether more dipoles lead to significantly better results.

\subsection{Optimal Performance of~$N$~Dipoles}
\label{eq:Ndips}

\begin{figure}
\centering
\includegraphics[width=\columnwidth]{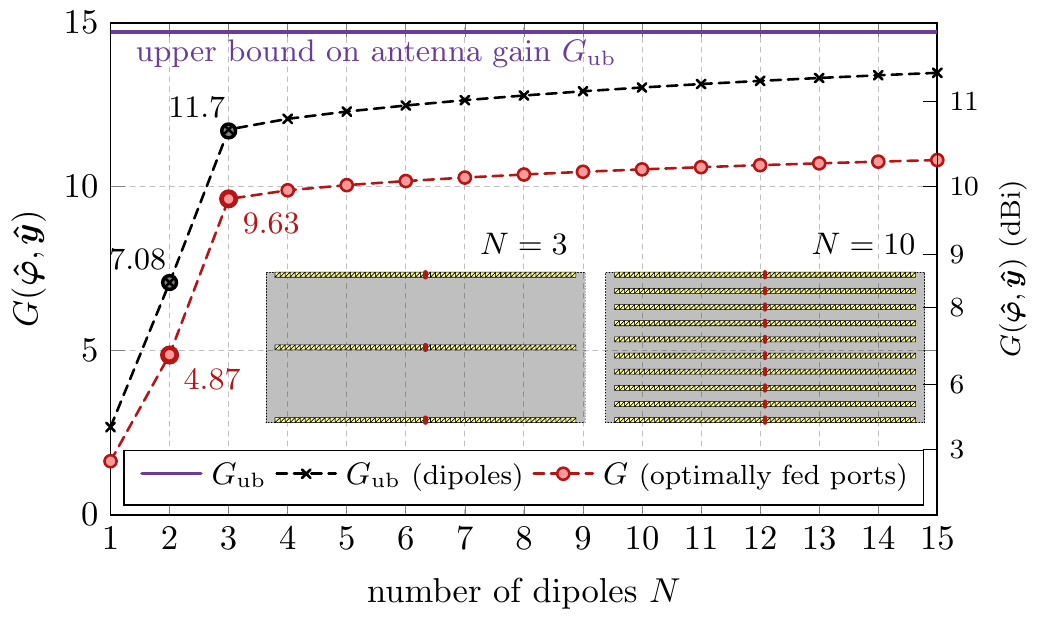}
\caption{Performance of dipoles placed in the design region of dimensions~$\ell \times \ell/2$. The length of the dipoles is~$0.945\ell$, and their width is~$\ell/60$. The dipoles are equidistantly spaced and made of copper. The upper bound for the entire design region is shown by the (constant) purple line and reaches~$\Gopt = 14.7$ (11.7\,dBi). When the optimal current is found only within the region delimited by the dipoles, the resulting gain is shown by the black cross marks (contoured by the black dashed line). The antenna gain for optimal excitation placed in the middle of each dipole is shown by the red circle marks (contoured by the red dashed line). Two examples of the dipole arrays are shown in the insets.}
\label{fig15}
\end{figure}

To enhance performance, the number of dipoles (the same geometry and excitation as before) is varied between~2 and~15, equidistantly occupying the design region~$\ell \times \ell/2$. The upper bound evaluated with the optimal current density occupying the entire region is depicted by a purple curve in Fig.~\ref{fig15}, while the upper bound corresponding to the region filled with dipoles only is shown by the black cross marks. The optimal delta-gap excitation of the dipoles (each fed in the middle) is shown by the red circular marks. The difference between the black and red curves is given by the fact that only $N$ feeders (placed in the middle of each dipole) cannot mimic the optimal current density. Despite this, the optimal excitation gives reasonable performance.

As expected, increasing the number of dipoles improves the gain performance with convergence to the upper bound on gain for the entire design region. Noticeably, three dipoles are capable of offering significantly better performance than two dipoles, \ie{}, $G=7.08$ (8.50\,dBi) vs $G=11.7$ (10.7\,dBi) for optimal current density, and $G=4.87$ (6.88\,dBi) vs $G=9.63$ (9.84\,dBi) for optimal feeding. Having four dipoles and more improves the performance just slightly, with remarkable diminishing returns occurring for $N>3$ dipoles, \cite[Ch.~5]{Collin_Zucker_AntennaTheoryPart1}.

\subsection{End-Fire Array of Three Dipoles}
\label{eq:3dipsPerf}

\begin{figure}
\centering
\includegraphics[width=\columnwidth]{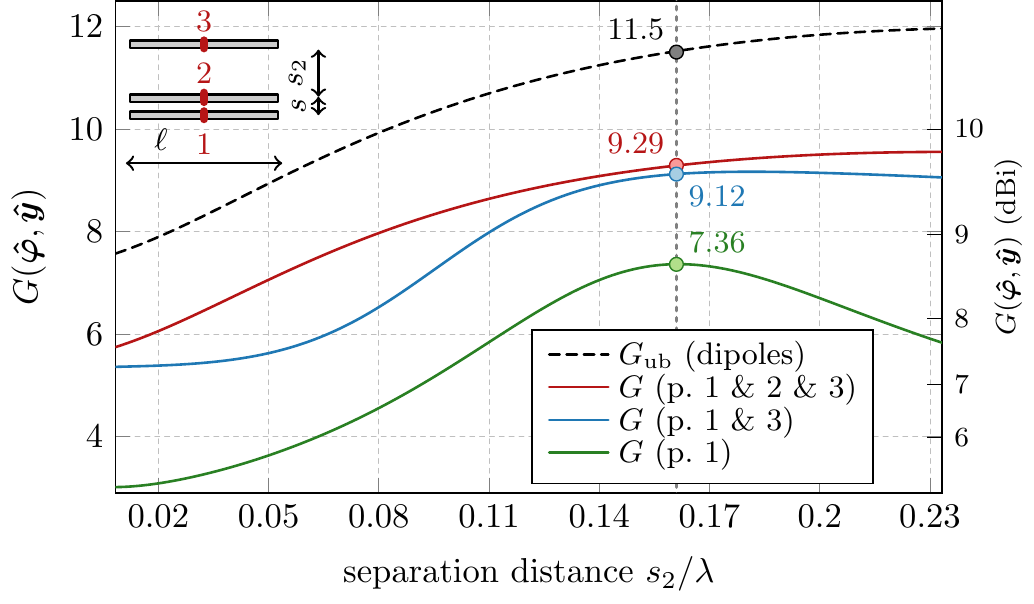}
\caption{Performance of end-fire array of three dipoles. The two-dipole arrangement is taken from Fig.~\ref{fig10}, \ie{}, center-center distance is~$s = 0.05\lambda$. The third dipole is added at center-center distance~$s_2$ (in the direction of radiation); see the inset for details and the numbering of the dipoles. The dashed black curve shows the upper bound for the entire region of three dipoles. The antenna gain for optimal excitation placed in the middle is shown by the red (all dipoles fed optimally), blue (1st and 3rd dipoles fed optimally, 2nd dipole is passive), and green curve (only 1st dipole fed, 2nd and 3rd dipoles are passive).}
\label{fig17}
\end{figure}

In this section, we elaborate on the case of three dipoles, starting from the design presented in Section~\ref{eq:2dipPerf} with two dipoles separated by~$s = 0.05\lambda$ and one feeder. The third dipole of the same length and material (copper) is added and separated by the distance~$s_2$, see inset of Fig.~\ref{fig17}. The same analysis as in Fig.~\ref{fig10} is repeated for three dipoles with distance~$s_2$ being swept.

The optimal current for the upper bound on the gain is evaluated via~\eqref{eq:GupFromCMAx} for the current supporting region restricted to the area of three dipoles. The upper bound~$\Gopt$ is shown by the black dashed line in Fig.~\ref{fig17} and increases with increasing distance~$s_2$. There is a saturation point close to $s_2/\lambda \approx 0.25$. The second study considers dipoles with optimal delta-gap feeds placed in the middle of each dipole. The current density has to obey constitutive relation~\eqref{eq:MoMeq} and the optimized variable is the excitation vector, which is entirely determined by the port voltages. Three cases are shown: all ports (see the inset of Fig.~\ref{fig17}), only two ports (1st and 3rd), and only one port (1st), are fed optimally. The curves in Fig.~\ref{fig17} represent these three scenarios, respectively, by the red, blue, and green colors. The port voltage can be chosen arbitrarily if only one port is excited. For other cases, there is a specific ratio between port voltages providing the maximal gain~\cite{Capeketal_OptimalityOfTARCAndRealizedGainForMultiPortAntennas, Ivrlac2010}. As expected, more control over the excitation provides higher gain. Considering all three cases, there is a distance, at approximately~$s_2 = 0.16\lambda$, where, irrespective of the number of ports, a high gain can be achieved. This separation distance is highlighted by the vertical dashed line and we will focus on it in the following section.

\subsection{Modal Analysis of the Three-Dipole Arrangement}
\label{eq:3dipsModal}

\begin{figure}
\centering
\includegraphics[width=\columnwidth]{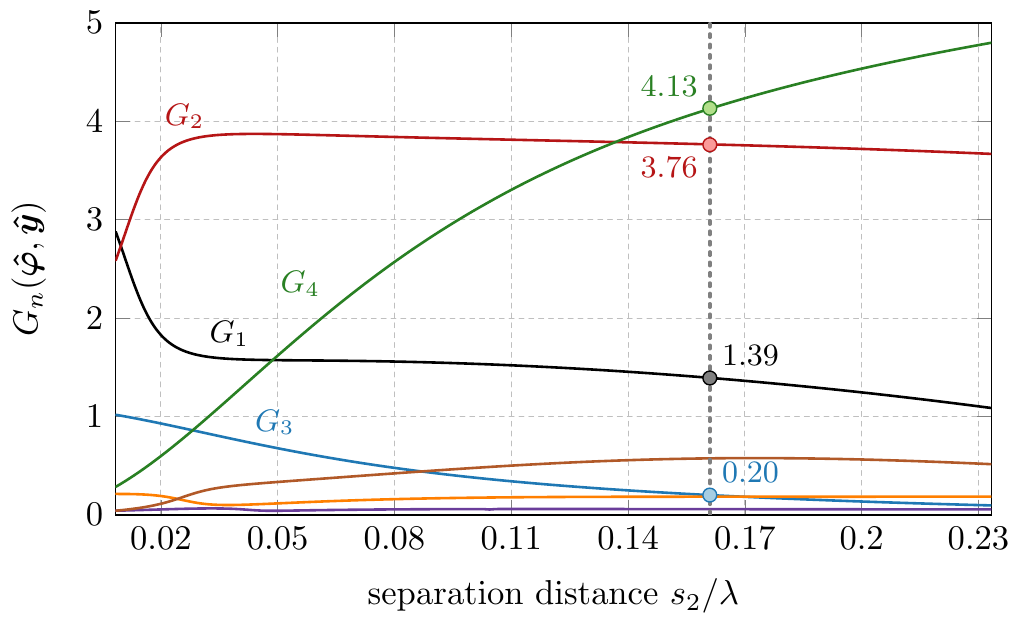}
\caption{Modal gains~$G_n$ of the current supporting region restricted by three dipoles of dimensions $0.945\ell \times w$ and separated by center-center distance~$s = 0.05\lambda$ (1st and 2nd dipole) and $s_2$ (2nd and 3rd dipole), see the inset in Fig.~\ref{fig17}. When summed up, the modal gains give the upper bound on gain~$\Gopt$, shown as the black dashed line in Fig.~\ref{fig17}.}
\label{fig18}
\end{figure}

The arrangement of three dipoles with varying separation distance~$s_2$ is studied from a modal point of view. The modal gains are shown in Fig.~\ref{fig18}. There are three modes mostly contributing to the upper bound on gain~$\Gopt$, namely~$G_1$ (black),~$G_2$ (red), and~$G_4$ (green). The curves are ordered with respect to the value of modal gain~$G_n$ for the smallest distance~$s_2$.

\begin{figure}
\centering
\includegraphics[width=\columnwidth]{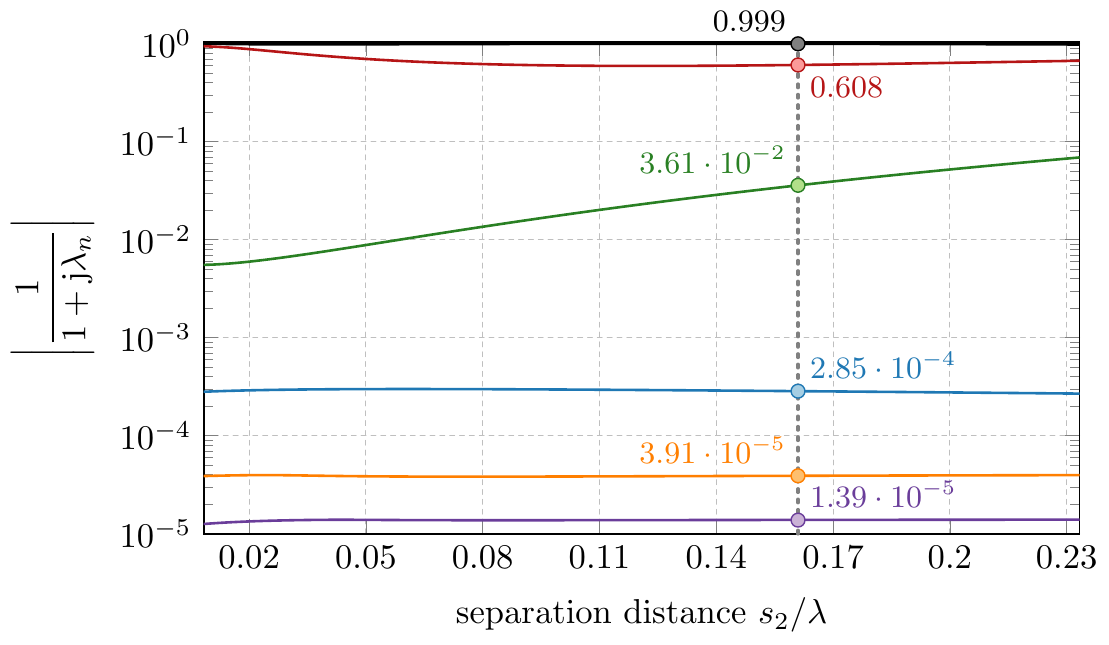}
\caption{Modal significance for the modes depicted in Fig.~\ref{fig18}. The same ordering and color-coding of the curves are used. Only three modes with modal gains~$G_1$, $G_2$, and~$G_4$, \cf{} Fig.~\ref{fig18}, have non-negligible modal significance, \ie{}, the ability to be excited.}
\label{fig19}
\end{figure}

The modal significance is shown in Fig.~\ref{fig19}. The modes with the highest modal gains are the only ones that can be sufficiently excited (significance is higher than 0.01). Therefore, realistically, we should expect the best performance in gain around $G_1+G_2+G_4 \approx 9.28$ (9.68\,dBi). This is confirmed by Fig.~\ref{fig17}, showing $G \approx 9.29$ when all three dipoles are fed optimally in their centers. However, in practice, the number of feeders should be minimized because of the complexity of the feeding network. Reactive loading can be used instead~\cite{Debard2023, Georgiadis2022}, albeit at the cost of losing degrees of freedom since the reactances must be passive.

Considering no reactive elements and lowering the number of feeding points, performance inevitably decreases, see Fig.~\ref{fig17}. A good trade-off is to utilize two feeders (for the 1st and 3rd dipoles). For distance~$s_2 = 0.16\lambda$, highlighted by the vertical dashed line, the performance is close to the case when all dipoles are fed ($G \approx 9.12$). Restricting the number of feeders to one (1st dipole), the performance drops to $G \approx 7.36$, still considerably above two dipoles with one feeder ($G \approx 5.38$, see Fig.~\ref{fig10}), but at the cost of the significantly larger overall height of the structure, \ie{}, $0.227\lambda$ (three dipoles) vs. $0.066 \lambda$ (two dipoles).

\begin{figure}
\centering
\includegraphics[width=\columnwidth]{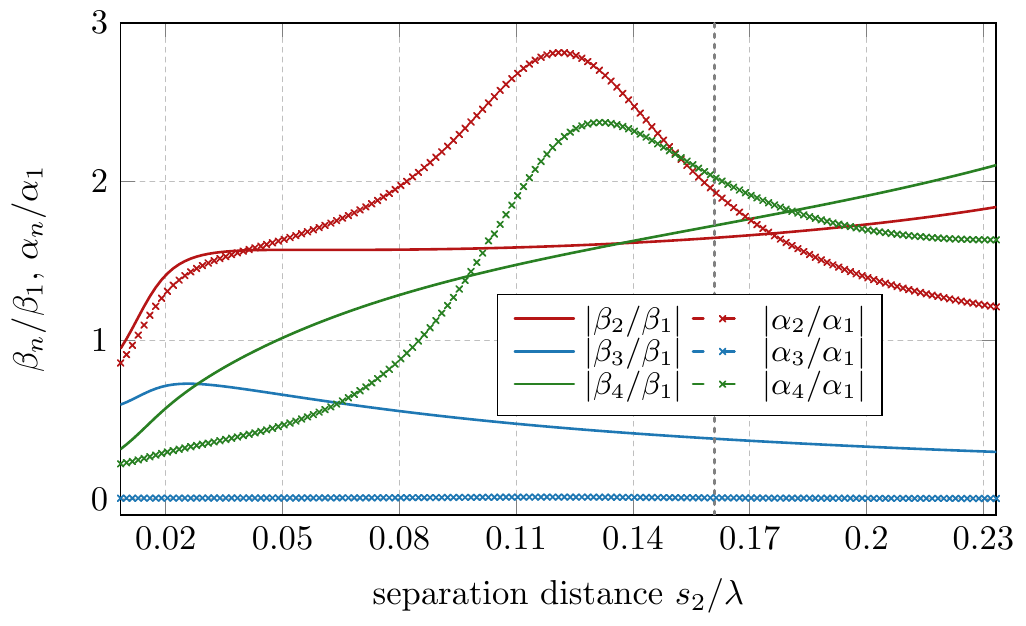}
\caption{Normalized optimal excitation coefficients~$\V{\beta}$, found via~\eqref{eq:betaCoef} and the normalized excitation coefficients~$\V{\alpha}$, evaluated via~\eqref{eq:excitCoef} for the 1st and 3rd dipoles fed optimally~\cite{Capeketal_OptimalityOfTARCAndRealizedGainForMultiPortAntennas}. The curves have the same meaning as in Fig.~\ref{fig13}, however, only the absolute values are shown to reduce the number of curves.}
\label{fig20}
\end{figure}

As a final note, the comparison of optimal excitation coefficients~$\V{\beta}$ required to optimally excite the first four modes with the highest gain ($G_1$--$G_4$) and the excitation coefficients~$\V{\alpha}$ realized by two feeders (1st and 3rd dipole) is shown in Fig.~\ref{fig20}. For the excitation, the optimal voltage is used as $v_1 = 1$\,V, and $v_3 = 0.79 \phase{-32^{\circ}}$. The results shown are ratios to the coefficients of the first mode, \ie{}, both $\V{\beta}$ and $\V{\alpha}$ vectors are renormalized to their first entries~$\beta_1$ and $\alpha_1$, respectively. It is seen that only the first, second, and fourth modes are nearly optimally excited (the solid lines are close to the dashed lines with cross marks), while the third mode is not excited at all ($\alpha_3 = 0$). Therefore, for better performance, more dipoles have to be added, and -- having Fig.~\ref{fig15} in mind -- a larger design region has to be used.

\section{Maximum Gain For Arbitrary Polarization}
\label{sec:highTotalGainX}

In Section~\ref{sec:maxGain}, the upper bound on antenna gain was found for fixed $\UV{e} = \UV{\vartheta}$ or $\UV{e} = \UV{\varphi}$ polarization. Here, the procedure is generalized for arbitrary polarization. This generally leads to higher antenna gain since polarization states are optimally combined to improve the upper bound.

A particular, yet interesting, case to start with is elliptical polarization which is fixed prior to the optimization. To get the upper bound, the far-field vector $\Fvec$ is composed of both principal polarizations as
\begin{equation}
    \label{eq:circPolar}
    \Fvec \left(\UV{e}, \UV{r}\right) = w_\vartheta \Fvec ( \UV{\vartheta}, \UV{r}) \pm \J w_\varphi \Fvec \left(\UV{\varphi}, \UV{r}\right),
\end{equation}
where the $\pm$ sign gives left/right-handed polarization  and $\vert w_\vartheta/w_\varphi\vert$ defines axial ratio~\cite{Balanis_Wiley_2005}. Matrix~$\Fvec$ is still of rank-1 and the procedure from Section~\ref{sec:maxGain} still applies.

When the polarization state is not known \textit{a priori}, the far-field matrix is composed of two polarizations~$\Fvec_\vartheta,\Fvec_\varphi$ and the optimal performance is evaluated from eigenvalue problem~\cite[(13)]{GustafssonCapek_MaximumGainEffAreaAndDirectivity}
\begin{equation}
    \label{eq:maxGainGEPeigFull}
\begin{bmatrix}
\Fvec_\vartheta^\herm &
\Fvec_\varphi^\herm
\end{bmatrix}
\begin{bmatrix}
\Fvec_\vartheta \\
\Fvec_\varphi
\end{bmatrix}
 \Iopt = \dfrac{\Gopt \ZVAC}{4\pi} \left(\Rmat + \Lmat \right) \Iopt.
\end{equation}

Following the same reasoning as in Section~\ref{sec:maxGain} and assuming
\begin{equation}
	\label{eq:GainIoptFull}
	\Iopt = \left(\Rmat + \Lmat\right)^{-1} \left( w_\vartheta \Fvec_\vartheta^\herm +
w_\varphi \Fvec_\varphi^\herm \right),
\end{equation}
where~$w_\vartheta, w_\varphi$ are arbitrary weights, the solution is given by a~$2\times 2$ eigenvalue problem
\begin{equation}
\label{eq:totGainMaxX}
    \sum\limits_n 
\begin{bmatrix}
F_{\vartheta n} F_{\vartheta n}^* & F_{\vartheta n} F_{\varphi n}^* \\
F_{\varphi n} F_{\varphi n}^* & F_{\varphi n} F_{\varphi n}^*
\end{bmatrix}
\begin{bmatrix}
w_\vartheta\\
w_\varphi
\end{bmatrix} = 
\dfrac{\Gopt \ZVAC}{4\pi} \begin{bmatrix}
w_\vartheta\\
w_\varphi
\end{bmatrix}.
\end{equation}
Eigenvectors of~\eqref{eq:totGainMaxX} are assumed to be normalized according to
\begin{equation}
\label{eq:eq:IoptNormX}
\begin{bmatrix}
w_\vartheta\\
w_\varphi
\end{bmatrix}^\herm
    \sum\limits_n 
\begin{bmatrix}
F_{\vartheta n} F_{\vartheta n}^* & F_{\vartheta n} F_{\varphi n}^* \\
F_{\varphi n} F_{\varphi n}^* & F_{\varphi n} F_{\varphi n}^*
\end{bmatrix}
\begin{bmatrix}
w_\vartheta\\
w_\varphi
\end{bmatrix} = 1.
\end{equation}

The upper bound on antenna gain is always highest when both polarizations are taken into account and their mixture is a subject of optimization, \ie{},~\eqref{eq:totGainMaxX} is employed. Nevertheless, for the principal directions, the upper bound is typically realized in the dominant polarization. This is the case of the example from Section~\ref{sec:Ex1} where the upper bound for $\UV{e} = \UV{\vartheta}$ polarization evaluated via~\eqref{eq:GupFromCMAx} is equal to the upper bound for the total polarization~\eqref{eq:totGainMaxX}.

\section{Effect of Self-Resonance}
\label{sec:selfResonance}

Self-resonance is an important additional constraint, especially for electrically small antennas~\cite{JelinekCapek_OptimalCurrentsOnArbitrarilyShapedSurfaces}. The problem reads
\begin{equation}
\begin{aligned}
	& \mathop \mathrm{maximize}\limits_\Ivec && G (\UV{e}, \UV{r}) \\
	& \mathrm{subject\,\,to} &&  \Ivec^{\herm} \Xmat \Ivec = 0.
\end{aligned}
\label{eq:maxGainSelfRes}
\end{equation}
The solution to problem~\eqref{eq:maxGainSelfRes}, with~$\Ivec$ being the optimization variable, was described in~\cite[Eq. (16)--(18)]{GustafssonCapek_MaximumGainEffAreaAndDirectivity} and reads
\begin{equation}
	\label{eq:maxgainSR}
	\Goptsr \left(\UV{e}, \UV{r}\right) = \min_x \max_{\Ivec} \dfrac{4 \pi}{\ZVAC} \dfrac{ \Ivec^\herm \M{K}^\herm (\UV{e}, \UV{r}) \M{K} (\UV{e}, \UV{r}) \Ivec}{\Ivec^\herm \left(\Rmat + \Lmat + x \Xmat \right) \Ivec}
\end{equation}
with parameter~$x$ being restricted to $x \in [-1/\max \left\{ \lambda_n \right\}, -1/\min \left\{\lambda_n \right\}]$, see \cite[App.~C]{GustafssonCapek_MaximumGainEffAreaAndDirectivity} for a detailed derivation. It should be noticed that for~$x = 0$, this problem reduces to that treated in Section~\ref{sec:maxGain}.

If current~$\Ivec$ is expanded into lossy characteristic modes via~\eqref{eq:excitCoef}, see Appendix~\ref{sec:AppA}, the maximum gain for self-resonant constraint can still be written as a sum of modal gains
\begin{equation}
    \label{eq:GnSR}
    \Goptsr = \sum_n \dfrac{G_n}{1 + x_\T{sr} \lambda_n},
\end{equation}
where the nominator contains modal gains~\eqref{eq:GupFromCMAx}, and the denominator contains correction coefficients~$x_\T{sr} \lambda_n$ to meet the self-resonance constraint $\Ivec^{\herm} \Xmat \Ivec = 0$.

The extra constraint of the self-resonant current in~\eqref{eq:maxGainSelfRes} is important for electrically small antennas where the current is otherwise highly capacitive/inductive. It is usually inactive for electrically large antennas. Considering the example from Section~\ref{sec:Ex1} (electrical size~$ka = 1.80$), upper bound~$\Gopt \approx 15.6$ (11.9\,dBi) is lowered to $\Goptsr \approx 14.4$ (11.6\,dBi) when a self-resonant current is desired.







\section{Conclusion}
\label{sec:Conclu}

The upper bound on antenna gain was expressed in a lossy characteristic mode basis as a simple sum of modal gains. The upper bound value can be easily evaluated in any software dealing with characteristic mode decomposition. The expression is valid for arbitrarily shaped obstacles, arbitrary material distribution, and arbitrary operational frequency. The upper bound can be established for an arbitrary polarization vector. An additional constraint on self-resonant optimal current was considered, mainly for electrically small antennas, lowering the bound. The cost in antenna Q-factor and radiation efficiency was determined, showing that the number of modes with acceptable modal parameters is low, lowering the performance in antenna gain as predicted by the upper bound.

The expansion into modes with zero cross-terms in antenna gain made it possible to compare the number of significant modes with the number of degrees of freedom predicted by the classical work of Harrington. Compared to his estimate on normal gain, the proposed bound can be significantly higher. This prevents the undesired possibility that the bound will be surpassed by an existing antenna design, a fact happening in the case of normal gain estimate. Nevertheless, a similar distinction of modes as normal and super-directive is proposed.

This work can be used in several ways. Apart from direct information regarding the upper bound on antenna performance, the associated modal parameters provide additional insight into the radiation mechanisms of an obstacle and its limits. This can be judged based on modal currents, modal significances, excitation coefficients, modal Q-factors, or modal radiation efficiencies and can help antenna design with further structural modifications of the antenna and its effective excitation. Finally, it was shown that the upper bound is unreachable with a realistic excitation. It can be, however, closely approached by a proper antenna design procedure.

\section*{Acknowledgement}
The authors would like to thank Richard W. Ziolkowski for an initial discussion about the validity of fundamental bounds resulting from his talk at IEEE APS/URSI 2022 in Denver.

\appendices

\section{Feasibility of the Upper Bound on Antenna Gain}
\label{sec:App0}

The inequality~$G \left(\UV{e}, \UV{r}\right) \leq \Gopt \left(\UV{e}, \UV{r}\right)$ can be rewritten into the basis of characteristic modes using~\eqref{eq:GmaxVsCMA},~\eqref{eq:excitCoef},~\eqref{eq:gain}, which results in
\begin{equation}
	\label{eq:gainAlphaBeta}
	\dfrac{4 \pi}{\ZVAC} \dfrac{\left\vert \sum\limits_n \alpha_n \Fvec (\UV{e}, \UV{r}) \Ivec_n \right\vert^2}{\V{\alpha}^\herm \V{\alpha}} \leq \dfrac{4 \pi }{\ZVAC} \sum_n \left\vert \Fvec (\UV{e}, \UV{r}) \Ivec_n \right\vert^2.
\end{equation}
Relation~\eqref{eq:gainAlphaBeta} holds for any vector of excitation coefficients~$\V{\alpha}$. Using abbreviation~$\M{F} (\UV{e}, \UV{r}) = \Fvec (\UV{e}, \UV{r}) \IVEC$, formula~\eqref{eq:gainAlphaBeta} is further rewritten as
\begin{equation}
    \left\vert \M{F} (\UV{e}, \UV{r}) \V{\alpha}  \right\vert \leq \left\vert \V{\alpha}  \right\vert \left\vert \M{F} (\UV{e}, \UV{r}) \right\vert,
\end{equation}
which is a generally valid Cauchy–Schwarz inequality.

\section{Self-resonant Bound As a Sum of Lossy Characteristic Modes}
\label{sec:AppA}

When current~$\Ivec$ is expanded into lossy characteristic modes~\eqref{eq:excitCoef}, relation~\eqref{eq:maxgainSR} reads
\begin{equation}
	\label{eq:gainCMAalpha2}
	\Goptsr \left(\UV{e}, \UV{r}\right) = \min_x \max_{\V{\alpha}} \dfrac{4 \pi}{\ZVAC} \dfrac{\V{\alpha}^\herm \M{F}^\herm (\UV{e}, \UV{r}) \M{F} (\UV{e}, \UV{r}) \V{\alpha}}{\V{\alpha}^\herm \IVEC^\herm \left(\Rmat + \Lmat + x \Xmat \right) \IVEC \V{\alpha}},
\end{equation}
where vector~$\M{F} (\UV{e}, \UV{r})$ aggregates characteristic far fields~$F_n (\UV{e}, \UV{r})$. Using further the orthogonal relations~\eqref{eq:CMAnorm} and~\eqref{eq:CMAnormX}, relation~\eqref{eq:gainCMAalpha2} can be written as
\begin{equation}
	\label{eq:gainCMAalpha3}
	\Goptsr \left(\UV{e}, \UV{r}\right) = \dfrac{4 \pi}{\ZVAC} \min_x \max_{\V{\alpha}} \dfrac{\V{\alpha}^\herm \M{F}^\herm (\UV{e}, \UV{r}) \M{F} (\UV{e}, \UV{r}) \V{\alpha}}{\V{\alpha}^\herm \left(\M{1} + x \V{\lambda} \right) \V{\alpha}},
\end{equation}
where $\M{1} \in \mathbb{R}^{N \times N}$ is the identity matrix and $\V{\lambda} \in \mathbb{R}^{N \times N}$ is the diagonal matrix of the characteristic numbers. The inner maximization problem is, as in Section~\ref{sec:maxGain}, resolved via
\begin{equation}
\label{eq:optAlpha}
    \V{\alpha} = \dfrac{\left(\M{1} + x \V{\lambda} \right)^{-1} \M{F}^\herm (\UV{e}, \UV{r})}{\vert \left(\M{1} + x \V{\lambda} \right)^{-1} \M{F}^\herm (\UV{e}, \UV{r}) \vert},
\end{equation}
where it is important to note that the inversion is made of a diagonal matrix. Substituting into~\eqref{eq:gainCMAalpha3} gives
\begin{equation}
	\label{eq:gainCMAalpha4}
	\Goptsr \left(\UV{e}, \UV{r}\right) = \min_x \left\{ \dfrac{4 \pi}{\ZVAC} \M{F} (\UV{e}, \UV{r}) \left(\M{1} + x \V{\lambda} \right)^{-1} \M{F}^\herm  (\UV{e}, \UV{r}) \right\},
\end{equation}
which is the result~\eqref{eq:GnSR} from Section~\ref{sec:selfResonance}.

\begin{figure}
\centering
\includegraphics[width=\columnwidth]{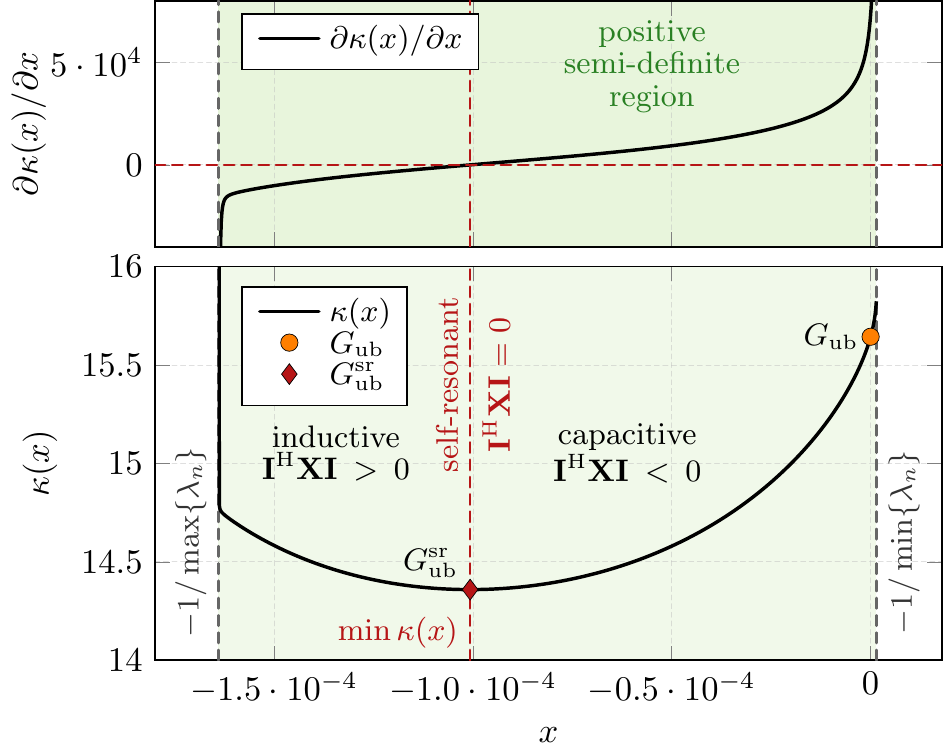}
\caption{Illustration of the convex optimization procedure to find an optimal mixture of lossy characteristic modes to maximize antenna gain for a self-resonant current. The data are shown for the example from Section~\ref{sec:Ex1}. The bottom pane shows the parameter $\kappa_(x)$, \eqref{eq:gainCMAalpha5}, to be minimized, and the top pane shows its derivative~\eqref{eq:derivative} used to accelerate the optimization procedure. The optimization domain is highlighted as the filled area and delimited by the smallest and largest characteristic eigenvalues. Two regions are classified as inductive (left) and capacitive (right).}
\label{figA}
\end{figure}

The function~$\kappa \left( x \right)$ to be minimized in~\eqref{eq:gainCMAalpha4} and its derivative can be further rewritten to
\begin{equation}
\label{eq:gainCMAalpha5}    
    \kappa \left( x \right) =  \dfrac{4\pi}{\ZVAC} \sum_n \dfrac{\left|  F_n (\UV{e}, \UV{r}) \right|^2}{1 + x \lambda_n}
\end{equation}
and
\begin{equation}
\label{eq:derivative}
    \dfrac{\partial \kappa (x)}{\partial x} = -  \dfrac{4\pi}{\ZVAC} \sum_n \dfrac{\lambda_n \left|  F_n (\UV{e}, \UV{r}) \right|^2}{\left( 1 + x \lambda_n \right)^2}
\end{equation}
and substituted into an optimization algorithm, \eg{}, a bisection rule, to find a proper halving interval, see Fig.~\ref{figA}. When a solution is found as~$x = x_\T{sr}$, we have~$\Goptsr = \min_x \kappa (x) = \kappa (x_\T{sr})$, and the optimal excitation coefficients are taken from~\eqref{eq:optAlpha} as~$\V{\beta}_\T{sr} = \V{\alpha} \left( x_\T{sr}\right)$.

Finally, once the solution exists, \ie{}, when one of the eigenvalues~$\lambda_n$ has an opposite sign than the others (which is always fulfilled for a good meshing), it can be shown that~$\Goptsr \leq \Gopt$. This is expected due to the additional constraint in~\eqref{eq:maxGainSelfRes}, which leads to additional minimization over parameter~$x$.

\begin{IEEEbiography}[{\includegraphics[width=1in,height=1.25in,clip,keepaspectratio]{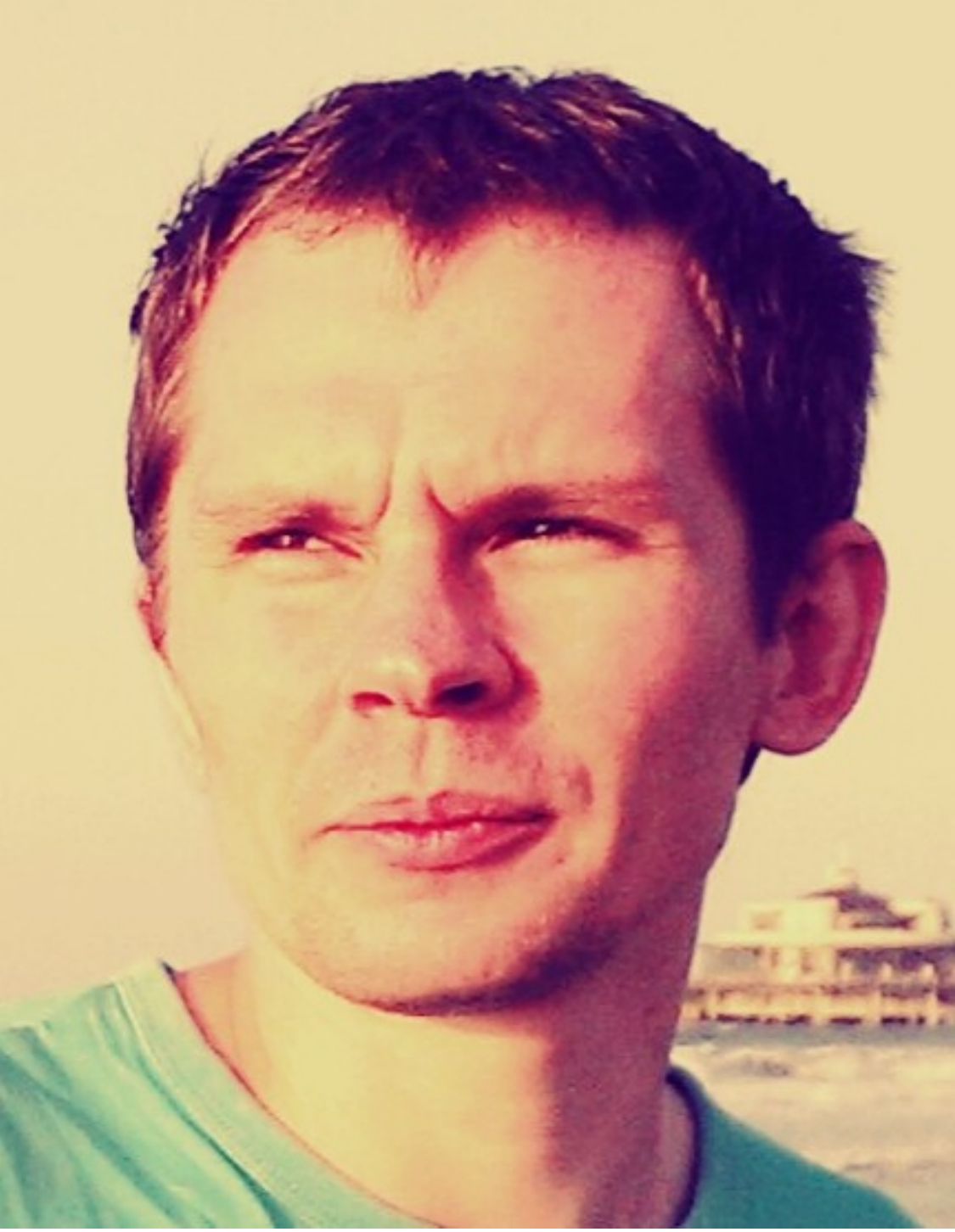}}]{Miloslav Capek}
(M'14, SM'17) received the M.Sc. degree in Electrical Engineering 2009, the Ph.D. degree in 2014, and was appointed Associate Professor in 2017, all from the Czech Technical University in Prague, Czech Republic.
	
He leads the development of the AToM (Antenna Toolbox for Matlab) package. His research interests are in the area of electromagnetic theory, electrically small antennas, numerical techniques, fractal geometry, and optimization. He authored or co-authored over 130~journal and conference papers.

Dr. Capek is the Associate Editor of IET Microwaves, Antennas \& Propagation. He was a regional delegate of EurAAP between 2015 and 2020. He received the IEEE Antennas and Propagation Edward E. Altshuler Prize Paper Award 2023.
\end{IEEEbiography}

\begin{IEEEbiography}[{\includegraphics[width=1in,height=1.25in,clip,keepaspectratio]{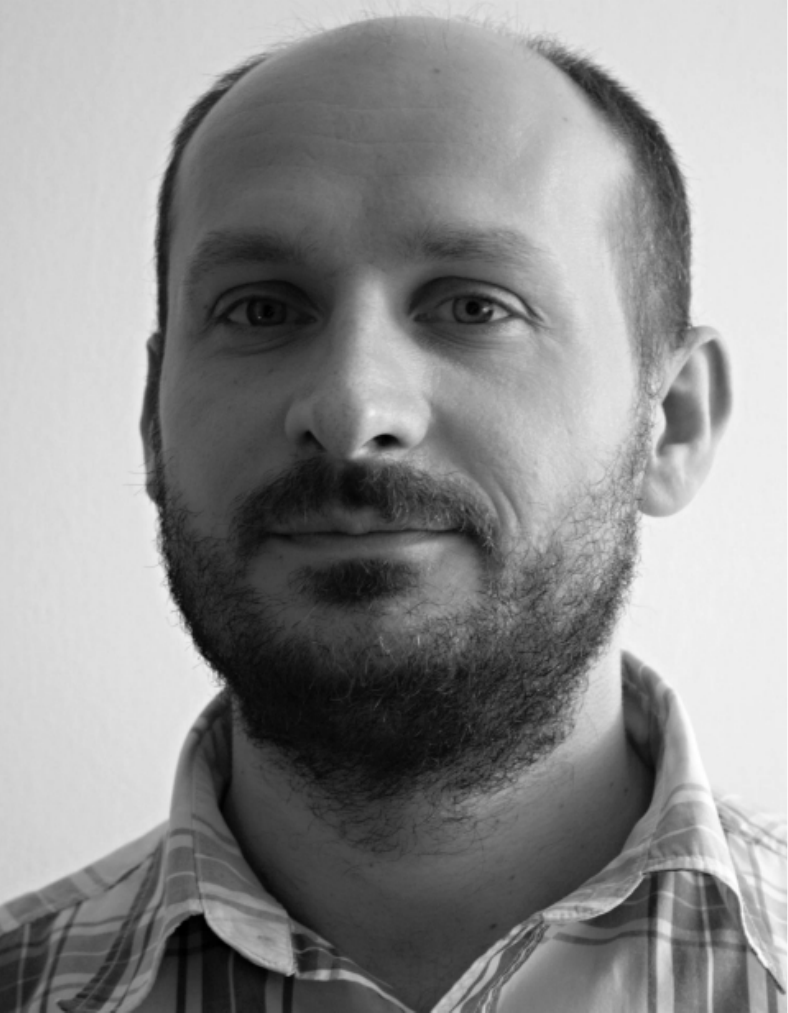}}]{Lukas Jelinek}
received his Ph.D. degree from the Czech Technical University in Prague, Czech Republic, in 2006. In 2015 he was appointed Associate Professor at the Department of Electromagnetic Field at the same university.

His research interests include wave propagation in complex media, electromagnetic field theory, metamaterials, numerical techniques, and optimization.
\end{IEEEbiography}

\end{document}